\documentclass[journal,onecolumn,12pt,twoside]{IEEEtran}

\usepackage{subfigure}
\usepackage{bm}
\usepackage{times}
\usepackage{epsf,epsfig}
\usepackage{float}
\usepackage{placeins}
\usepackage{psfrag}
\usepackage{amsmath}
\usepackage{bm}
\usepackage{amssymb}
\usepackage{graphicx}
\usepackage{epstopdf}
\usepackage{multirow}
\usepackage{enumerate}
\usepackage{color}
\usepackage{mathrsfs}
\usepackage{setspace}
\usepackage{diagbox}
\usepackage{enumitem}
\usepackage[table]{xcolor}

\makeatletter
\newif\if@restonecol
\makeatother

\usepackage[linesnumbered,ruled,vlined]{algorithm2e}
\usepackage{algpseudocode}
\ifCLASSINFOpdf
\else
\fi

\begin{document}
\title{Spatially Coupled PLDPC-Hadamard Convolutional Codes }

\author{
        Peng W.  Zhang, 
        Francis C.M. Lau,~\IEEEmembership{Fellow,~IEEE,} 
        and~Chiu-W. Sham,~\IEEEmembership{Senior Member,~IEEE}
\thanks{P.~W. Zhang was  with the Future Wireless Networks and IoT Focusing Area,
         Department
of Electronic and Information Engineering, The Hong Kong Polytechnic University, Hong Kong SAR, China. He is now with Huawei Technologies Ltd., Chengdu, China.}
      \thanks{  F.~C.~M. Lau is  with the Future Wireless Networks and IoT Focusing Area,
         Department
of Electronic and Information Engineering, The Hong Kong Polytechnic University, Hong Kong SAR, China.  (e-mail: francis-cm.lau@polyu.edu.hk).}
\thanks{C.-W. Sham is with the Department of Computer Science,  The University of Auckland,
New Zealand (e-mail: b.sham@auckland.ac.nz).}
\thanks{  The work described in this paper was partially supported by the RGC Research Impact Fund from the Hong Kong SAR, China (Project No. R5013-19).}
}

\maketitle
\begin{abstract}
We  propose a new type of ultimate-Shannon-limit-approaching codes called
spatially coupled protograph-based low-density parity-check Hadamard convolutional codes (SC-PLDPCH-CCs), which
are constructed by spatially coupling PLDPC-Hadamard block codes.
We develop an efficient decoding algorithm that combines pipeline decoding and layered scheduling for the decoding
of SC-PLDPCH-CCs, {\color{black}and analyze the latency and complexity of the decoder}.
To estimate the decoding thresholds of SC-PLDPCH-CCs,
we first propose a layered protograph extrinsic information transfer (PEXIT) algorithm to evaluate the thresholds of
  spatially coupled PLDPC-Hadamard terminated codes (SC-PLDPCH-TDCs)
  with a moderate coupling length.
With the use of the proposed layered PEXIT method, we develop a genetic algorithm to find good SC-PLDPCH-TDCs in a systematic way.
Then 
 we extend the coupling length of these SC-PLDPCH-TDCs to form good SC-PLDPCH-CCs.
Results show that our constructed SC-PLDPCH-CCs can achieve comparable thresholds to the block code counterparts.
Simulations illustrate the superiority
of the SC-PLDPCH-CCs over the block code counterparts and other state-of-the-art low-rate codes in terms of error performance.
For the rate-$0.00295$ SC-PLDPCH-CC, a bit error rate of {\color{black}$10^{-5}$ is achieved at $E_b/N_0 = -1.465$ dB, which is only $0.125$ dB} from the ultimate Shannon limit.
\end{abstract}


\IEEEpeerreviewmaketitle

\section{Introduction}
{\color{black}Binary} low-density parity-check (LDPC) codes were first proposed in 1963 \cite{Gallager1963}, whose sparse parity-check matrix containing ``$0$'' and ``$1$'' is randomly constructed, and can be graphically represented by a Tanner graph \cite{Tanner1981}.
A Tanner graph consists of check nodes (CNs), variable nodes (VNs) and the connections between CNs and VNs.
After receiving the channel observations, extrinsic information along the connections is iteratively exchanged and calculated at CN processors and VN processors to realize belief propagation (BP) decoding.
In \cite{Richardson2001}, a density evolution method is proposed to
evaluate the probability density function (PDF) of the extrinsic information
and hence to optimize the degree distributions of LDPC codes.
Subsequently in \cite{Brink2004},  an extrinsic information transfer (EXIT) chart technique
is proposed to optimize LDPC codes by calculating the mutual information (MI) of the extrinsic information.
Through both methods, good-performing LDPC codes are constructed
to work close to the Shannon limit under binary erasure channels (BECs),  binary symmetric channels (BSCs)
 and   additive-white-Gaussian-noise (AWGN) channels \cite{Richardson2001,Brink2004,Sharon2006b}.
Yet, most LDPC codes are designed to achieve good decoding performance at a bit-energy-to-noise-power-spectral-density ratio ($E_b/N_0$) greater than $0$ dB.
To approach the ultimate Shannon limit, i.e., $E_b/N_0 = -1.59$ dB \cite{Costello2007}, other low-rate  codes are designed.
Potential application scenarios of such codes include multiple access wireless systems (e.g., interleave-division multiple-access \cite{Li2004,Li2006} with a huge number of non-orthogonal users) and deep space communications.
In \cite{Li2003} and \cite{Leung2006}, very low-rate
 turbo Hadamard codes and zigzag Hadamard codes are proposed.
However,
their decoders require the use of serial decoding \cite{Jiang2018}, \cite{Jiang2019}, \cite{Jiang2020} and cannot make use of
 parallel decoding as in LDPC decoders. The error performances
 of turbo Hadamard codes and zigzag Hadamard codes are also not as good as
those of the LDPC-Hadamard codes that {\color{black} have been} subsequently proposed \cite{Yue2007}.

In the Tanner graph of an LDPC code, the edges connected to a VN
form a repeat code, whereas the edges connected to a CN
form a single-parity-check (SPC) code.
When the repeat codes and/or SPC codes are replaced with other block codes,
a Tanner graph of a generalized LDPC is formed.
By replacing SPC codes with Hadamard codes, LDPC-Hadamard codes 
(LDPC-HCs) have been proposed
 and  their degree distributions have been optimized with the EXIT chart method \cite{Yue2007}.
The 
LDPC-HCs not only have thresholds lower than $-1.34$ dB, but also achieve excellent decoding performance at a $E_b/N_0$ lower than $-1.17$ dB.
On the other hand, the EXIT chart method cannot effectively analyze degree distributions containing degree-$1$ and/or punctured VNs.
The parity-check matrix derived by the progressive-edge-growth (PEG) method \cite{Hu2005} according to the degree distributions does not contain any structure
and therefore does not facilitate linear encoding and parallel decoding.
In  \cite{zhang2021}, a class of structured 
LDPC-HCs
called protograph-based LDPC-HCs 
(PLDPC-HCs) have been
proposed. 

Protograph-based LDPC (PLDPC) codes can be described by a small protomatrix or protograph \cite{Thorpe2003}.
Using the copy-and-permute operations to lift the protomatrix or protograph,
the derived matrix or lifted graph can easily have a quasi-cyclic structure which is conducive to the hardware implementations of encoders and decoders  \cite{Fossorier2004}.
PLDPC-HCs \cite{zhang2021} also inherit such advantages from PLDPC codes.
By adding an appropriate amount of degree-$1$ Hadamard variable nodes (D1H-VNs) to the check nodes in the protograph of a PLDPC code, the SPC constraints are converted into Hadamard constraints and the protograph of a  
PLDPC-HC is obtained.
The protograph-based EXIT (PEXIT) chart method for PLDPC codes \cite{Liva2007} has
further been modified for analyzing 
PLDPC-HCs \cite{zhang2021}.
The modified method not only can produce multiple extrinsic mutual information
from the symbol-by-symbol maximum-a-posteriori-probability (symbol-MAP) Hadamard decoder, but also is applicable to analyzing protographs with degree-$1$ and/or punctured VNs.
The structured 
PLDPC-HCs
can achieve good thresholds and comparable error performance as the traditional 
LDPC-HCs.
In \cite{Zhang2021layer},  a layered decoding algorithm has been proposed
 to double the convergence rate  compared with standard decoding.

While 
PLDPC-HCs
 are already working very close to the ultimate Shannon limit,
further improvement is possible. In \cite{Iyengar2012}, it has been shown that
LDPC convolutional codes (LDPC-CCs) can achieve convolutional gains over their
block-code counterparts.
LDPC-CCs were first proposed in \cite{Jimenez1999} and characterized by the degree distributions of the underlying LDPC block codes (LDPC-BCs).
In addition, spatially coupled LDPC (SC-LDPC) codes are constructed by coupling
 a number of LDPC-BCs.
As the number of coupled LDPC-BCs tends to infinity, spatially coupled LDPC convolutional codes (SC-LDPC-CCs) are obtained.
In \cite{Kudekar2011} and \cite{Kudekar2013}, SC-LDPC-CCs have been shown to
 achieve capacity over binary memoryless symmetric channels under BP decoding.
In \cite{Stinner2016},  SC-LDPC-CCs have been constructed from the perspective of protographs, namely SC-PLDPC-CCs.
Through the edge-spreading procedure on protomatrix,  the threshold, convergence behavior and error performance of SC-PLDPC ensembles have also been systematically investigated.
{\color{black}In particular, PEXIT algorithms have been applied to analyze
 binary \cite{Battaglioni2017} and $q$-ary SC-LDPC codes \cite{Wei2016}.}
Based on the aforementioned results, spatially coupled PLDPC-Hadamard convolutional codes (SC-PLDPCH-CCs), which are to be investigated in this paper, have the potential to provide extra gains over their block-code counterparts.
A genetic algorithm (GA) will also be proposed to optimize the design of SC-PLDPCH-CCs.

GA is an optimization algorithm that simulates the evolution of nature, and is widely used in music generation, genetic synthesis and VLSI technology \cite{Srinivas1994}.
In the arena of channel codes, GA has been applied to adjust the code rate of turbo codes without puncturing \cite{Hebbes2005};
to construct polar codes that 
reduce the decoding complexity while maintaining the same decoding performance
\cite{Elkelesh2019b};
together with bit-error-rate (BER) simulations to optimize error performance of short length LDPC codes over both AWGN and Rayleigh fading channels  \cite{Elkelesh2019};
together with density evolution to optimize the degree distributions of the SC-LDPC codes over BEC channels
\cite{Koganei2016}.
 {\color{black}
 GA first forms an offspring generation from the parent generation, 
 and performs operations such as crossover, mutation, and selection on the offspring generation.
A similar technique called differential evolution algorithm directly uses the parent generation to perform these operations to achieve the evolution. 
Differential evolution based optimization methods have been applied to design LDPC and SC-LDPC codes \cite{Liao2021DAAODB,Liao2021CSCLDPC}.
Compared with the differential evolution algorithm, the genetic algorithm has a stronger global search ability but requires more complex procedures.
  }

In this paper we propose a new type of ultimate-Shannon-limit-approaching codes, namely
spatially coupled PLDPC-Hadamard convolutional codes (SC-PLDPCH-CCs),
and  {\color{black} we conduct} an in-depth investigation into the proposed codes.
Our main contributions are as follows.
\begin{itemize}
\item[1)] We propose a new type of ultimate-Shannon-limit-approaching codes, namely
spatially coupled PLDPC-Hadamard convolutional codes (SC-PLDPCH-CCs),
which are constructed by spatially coupling PLDPC-Hadamard block codes.

\item[2)] We describe the encoding method of SC-PLDPCH-CCs. We also develop an efficient decoding algorithm, i.e., a pipeline decoding strategy combined with layered scheduling, for the decoding of SC-PLDPCH-CCs, {\color{black} and analyze its latency and complexity}.

\item[3)]
Using the original PEXIT method in \cite{zhang2021}, we show that the thresholds of different spatially coupled PLDPC-Hadamard terminated codes (SC-PLDPCH-TDCs) are distinguishable and improves as the coupling length increases.
To improve the convergence rate of the original PEXIT method, we propose a layered PEXIT  
algorithm to efficiently evaluate the threshold of SC-PLDPCH-TDCs with a
given  coupling length.
The thresholds of SC-PLDPCH-TDCs with large coupling lengths are then used as  estimates for thresholds of SC-PLDPCH-CCs.

\item[4)]
We propose a GA
 to systematically search for SC-PLDPCH-TDCs having good thresholds.
Based on the same set of split protomatrices for good SC-PLDPCH-TDCs, we extend the coupling length to construct the convolutional codes, i.e., SC-PLDPCH-CCs.

\item[5)] We have found SC-PLDPCH-CCs with comparable thresholds to the underlying  PLDPC-Hadamard block codes (PLDPCH-BCs).
Simulation results show that SC-PLDPCH-CCs outperform their 
PLDPCH-BC 
counterparts {\color{black}and other state-of-the-art low-rate codes} in terms of bit error performance.
For the rate-$0.00295$ SC-PLDPCH-CC, a  
{\color{black}BER of $10^{-5}$ is achieved at
$E_b/N_0 = -1.465$ dB.}
\end{itemize}

Section \ref{sect:contributions} of this paper highlights the main differences/improvements
of the current work compared with the previously studied 
PLDPC-HCs.
Section \ref{sect:background}  reviews the structures of related block codes and spatially coupled codes.
Section \ref{sect:SC-PLDPCH-CC} introduces the structure and encoding process  of SC-PLDPCH-CCs;
 describes a pipeline decoding strategy combined with layered scheduling
for decoding SC-PLDPCH-CCs, and analyzes its latency and complexity.
Also, a layered PEXIT chart method is proposed to  evaluate the threshold of SC-PLDPCH-TDCs/SC-PLDPCH-CCs efficiently and a GA is proposed to optimize protomatrices for SC-PLDPCH-TDCs/SC-PLDPCH-CCs.
Section \ref{sect:sim_re} presents the thresholds and optimized protomatrices of SC-PLDPCH-CCs with different code rates. It also compares the  simulated BER results of the SC-PLDPCH-CCs with those
of the underlying protograph-based PLDPC-Hadamard block codes (PLDPCH-BCs)
{\color{black}and other state-of-the-art low-rate codes}.
Finally, Section \ref{sect:conclusion} presents some concluding remarks.

{\color{black}
\section{Main differences/improvements between this work and previous works} \label{sect:contributions}
The major differences/improvements between the current work and the previously studied 
PLDPC-HCs
 \cite{zhang2021,Zhang2021layer} are as follows.
\begin{enumerate}
\item
The  
PLDPC-HCs 
 in  \cite{zhang2021,Zhang2021layer} and
the codes being investigated in this work are constructed based on protographs with the SPC check nodes (SPC-CNs) replaced by Hadamard constraints.
However, the 
PLDPC-HCs 
studied in  \cite{zhang2021,Zhang2021layer} are block codes
while
the codes being investigated in this work are formed by spatially coupling these block codes.
\item Though PEXIT algorithms have been applied for analyzing the codes in \cite{zhang2021} and in this work, 
 a layered PEXIT algorithm is proposed in this work to improve
the convergence rate of the original PEXIT algorithm in \cite{zhang2021}.
Also, a more efficient way of evaluating the
extrinsic 
MI 
of the Hadamard 
CNs
is applied
in the algorithm here compared with  \cite{zhang2021}.
\item A 
GA is proposed in this work to search for spatially-coupled 
PLDPC-HCs
with good theoretical thresholds.
\item
We estimate the thresholds of SC-PLDPCH-CCs based on the thresholds of SC-PLDPCH-TDCs with large coupling lengths.
\item The PLDPC-Hadamard sub-decoders in \cite{Zhang2021layer} and
the processors  in this work are of similar structures. All the them apply
layered decoding algorithms for decoding the 
PLDPC-HCs.
However, the structure of the PLDPC-Hadamard sub-decoders in \cite{Zhang2021layer} is
more or less fixed for a given code design; and the number of decoding iterations affects
the error performance of the code, the decoding latency and throughput.
The pipeline decoder described in this work consists of a series of processors, the number of which can vary;
and the number of processors in the decoder affects the error performance  of the code, the decoding latency,
the hardware complexity and throughput.
\item The SC-PLDPCH-CCs in this work can achieve a better error performance
and throughput than the 
PLDPCH-BCs
\cite{zhang2021}
with a higher hardware requirement.
\end{enumerate}
}

\section{Background} \label{sect:background}
\subsection{PLDPC Block Codes}\label{sect:ldpc-bc}
A protograph consists of a set of $m$ SPC-CNs,
a set of $n$ ($n > m$) protograph variable nodes (P-VNs),
and a set of edges connecting the SPC-CNs to the P-VNs  \cite{Thorpe2003}. 
The corresponding protomatrix
can be denoted by $\bm{B}=\{b(i,j)\}$
in which each row in $\bm{B}$ corresponds to a SPC-CN;
each column corresponds to a P-VN;
and $b(i,j)$ corresponds to the number of edges connecting
 the $i$-th SPC-CN and the $j$-th P-VN.
The code rate of a PLDPC block code equals
$R_{\rm PLDPC-BC} = 1- \frac{m}{n}$.
A two-step lifting method 
(with factors $z_1$ and $z_2$) can be used to lift the protomatrix and hence to construct the parity-check matrix
of a PLDPC code \cite{Wang2013}.
After the first lifting, all entries in the lifted matrix {\color{black} are} either ``0'' or ``1''.
The second lifting procedure aims to construct a parity-check matrix with a quasi-cyclic structure
so as to simplify encoder and decoder designs.

\subsection{LDPC-Hadamard Codes and PLDPC-Hadamard Codes}\label{subsec:PLDPCH}
A Hadamard code with an order $r$ has a code length of $q = 2^r$
and can be obtained by a Hadamard matrix $\bm{H}_q$ of size $q \times q$.
$\bm{H}_q$ can be  recursively generated using \cite{Li2003, Yue2007}
$\pm {\bm{H}_q} = \pm \{\bm{h}_{j}\} = \left[ {\begin{array}{*{20}{c}}
{ \pm {\bm{H}_{q/2}}}&{ \pm {\bm{H}_{q/2}}}\\
{ \pm {\bm{H}_{q/2}}}&{ \mp {\bm{H}_{q/2}}}
\end{array}} \right]$
where $\pm \bm{H}_1 = [\pm 1]$;
$\pm\bm{h}_j = [\pm h_{0,j}\ \pm h_{1,j}\ \cdots\ \pm h_{2^r-1,j}]^T$ ($j = 0,1,\ldots q-1$) represents the
$j$-th column of $\pm {\bm{H}_q} $ and is a Hadamard codeword.
Suppose $+1$ is mapped to bit ``$0$'' and $-1$ to bit ``$1$''.
It has been shown that when the Hadamard order $r$ is an even number,
the $0$th, $1$st, $2$nd, $\ldots$, $2^{k-1}$-th, $\ldots$, $2^{r-1}$-th, and $(2^r-1)$-th
code bits (total $r+2$ bits) in each codeword $\pm\bm{h}_j$ satisfy the SPC constraint together, i.e., \cite{Yue2007,zhang2021}
\begin{eqnarray}
\label{eq:spc}
&& \makebox[-1.5cm]{}  [ \pm h_{0,j}  \oplus \pm h_{1,j} \oplus  \cdots\oplus  {\ \pm h_{2^{k-1},j}} \cr
&& \oplus \cdots  \oplus  {\ \pm h_{2^{r-1},j}} ]\   
\oplus\ \pm h_{2^r-1,j}   = 0,
\end{eqnarray}
where the symbol $\oplus$  represents the XOR operator.
%
When the SPC codes of an LDPC code are replaced with Hadamard block codes with additional Hadamard parity-check bits, 
an LDPC-Hadamard block code (LDPCH-BC) is formed  \cite{Yue2007}.
LDPCH-BCs can be constructed by lifting protographs in which
the SPC-CNs have been replaced with Hadamard check nodes (H-CNs) and additional parity-check bits.
Such 
LDPCH-BCs are called
PLDPCH-BCs \cite{zhang2021}.
We assume that each SPC-CN  in the original protograph connects $d$ P-VNs.
The corresponding $d$ bit values can be used as information bits and input to a systematic Hadamard encoder with an order $r$,
where $r = d - 2$. When $r$ is an even number, these $d=r+2$ inputs correspond to those entries
in \eqref{eq:spc},
based on which the remaining $2^r-d$ Hadamard parity-check bits 
(represented by 
D1H-VNs in the protograph)
are found.
{\footnote{When $r$ is an odd number, the Hadamard code bits do not satisfy \eqref{eq:spc}.
Thus,
a non-systematic Hadamard encoding scheme is used and $2^r-2$ Hadamard parity-check bits are generated
 \cite{zhang2021}. Here, we focus our analysis on cases where $r$ is even.}
Fig. \ref{fig:PLDPCH_1}(a) illustrates the protograph of a PLDPCH-BC consisting of $m=3$ H-CNs and $n=4$ P-VNs. 
Using a two-step lifting process with factors $z_1$ and $z_2$,
a PLDPCH-BC can be constructed 
with an information length
of $(n - m) z_1 z_2$.
When $r$ is even, the overall code length equals
$n z_1 z_2 + m(2^r - r - 2) z_1z_2$ \cite{zhang2021}
and the code rate of the PLDPCH-BC is given by
$R^{\rm even}_{\rm PLDPCH-BC} = \frac{{n - m}}{{n + m\left( {{2^r} - r - 2} \right)}}$.


\subsection{Spatially Coupled LDPC Codes}\label{sc-ldpc-cc}
\subsubsection{LDPC convolutional codes}
The parity-check matrix $\bm{H}_{\rm CC}$ of an 
LDPC-CC \cite{Jimenez1999}
is semi-infinite and structurally repeated. $\bm{H}_{\rm CC}$
can be written as in \eqref{eq:H_cc},
where each ${\bm{H}_i}(t)$ ($i = 0, 1, \ldots, m_s$) is {\color{black} an} $M \times N$ component parity-check matrix, $t$ denotes
the time index, and $m_s$ is the {syndrome former memory}.
Each codeword $\bm{c}$ satisfies $\bm{c} {\bm{H}_{\rm CC}}^T = \bm{0}$, where $\bm{0}$ is a semi-infinite zero vector.

\begin{figure*}
\begin{equation}
{\bm{H}_{\rm CC}} = \left[ {\begin{array}{*{20}{c}}
{{\bm{H}_0}\left( 1 \right)}&{}&{}&{}&{}\\
{{\bm{H}_1}\left( 1 \right)}&{{\bm{H}_0}\left( 2 \right)}&{}&{}&{}\\
 \vdots &{{\bm{H}_1}\left( 2 \right)}& \ddots &{}&{}\\
{{\bm{H}_{{m_s}}}\left( 1 \right)}& \vdots & \ddots &{{\bm{H}_0}\left( t \right)}&{}\\
{}&{{\bm{H}_{{m_s}}}\left( 2 \right)}& \ddots &{{\bm{H}_1}\left( t \right)}& \ddots \\
{}&{}& \ddots & \vdots & \ddots \\
{}&{}&{}&{{\bm{H}_{{m_s}}}\left( t \right)}& \ddots \\
{}&{}&{}&{}& \ddots
\end{array}} \right] 
\label{eq:H_cc} 
\end{equation}
\begin{equation}\label{eq:td}
\bm{B}_{\rm SC-PLDPC-TDC} = \overbrace {\left[ {\begin{array}{*{20}{c}}
{{\bm{B}_0}}&{}&{}&{}\\
{{\bm{B}_1}}&{{\bm{B}_0}}&{}&{}\\
 \vdots &{{\bm{B}_1}}& \ddots &{}\\
{{\bm{B}_W}}& \vdots & \ddots &{{\bm{B}_0}}\\
{}&{{\bm{B}_W}}& \ddots &{{\bm{B}_1}}\\
{}&{}& \ddots & \vdots \\
{}&{}&{}&{{\bm{B}_W}}
\end{array}} \right]}^{nL}\left.\!\!\!\!\!\!\!\!\!\! \begin{array}{l}
\\
\\
\\
\\
\\
\\
\\
\\
\end{array} \right\}{^{m(L + W)}}.
\end{equation}
\begin{equation}
\bm{B}_{\rm SC-PLDPC-TBC} = \overbrace {\left[ {\begin{array}{*{20}{c}}
{{\bm{B}_0}}&{}&{}&{}&{}&{{\bm{B}_W}}& \cdots &{{\bm{B}_1}}\\
{{\bm{B}_1}}&{{\bm{B}_0}}&{}&{}&{}&{}& \ddots & \vdots \\
 \vdots &{{\bm{B}_1}}&{{\bm{B}_0}}&{}&{}&{}&{}&{{\bm{B}_W}}\\
{{\bm{B}_W}}& \vdots &{{\bm{B}_1}}& \ddots &{}&{}&{}&{}\\
{}&{{\bm{B}_W}}& \vdots & \ddots &{{\bm{B}_0}}&{}&{}&{}\\
{}&{}&{{\bm{B}_W}}& \ddots &{{\bm{B}_1}}&{{\bm{B}_0}}&{}&{}\\
{}&{}&{}& \ddots & \vdots & \ddots &{{\bm{B}_0}}&{}\\
{}&{}&{}&{}&{{\bm{B}_W}}& \cdots &{{\bm{B}_1}}&{{\bm{B}_0}}
\end{array}} \right]}^{nL}\left.\!\!\!\!\!\!\!\! \begin{array}{l}
\\
\\
\\
\\
\\
\\
\\
\\
\\
\\
\\
\end{array} \right\}{^{mL}}
\label{eq:tbc}
\end{equation}
\begin{equation}\label{eq:cc_mat}
{\bm{B}_{\rm SC-PLDPC-CC}} = \left[ {\begin{array}{*{20}{c}}
{{\bm{B}_0}}&{}&{}&{}&{}\\
{{\bm{B}_1}}&{{\bm{B}_0}}&{}&{}&{}\\
 \vdots &{{\bm{B}_1}}& \ddots &{}&{}\\
{{\bm{B}_W}}& \vdots & \ddots &{{\bm{B}_0}}&{}\\
{}&{{\bm{B}_W}}& \ddots &{{\bm{B}_1}}& \ddots \\
{}&{}& \ddots & \vdots & \ddots \\
{}&{}&{}&{{\bm{B}_W}}& \ddots \\
{}&{}&{}&{}& \ddots
\end{array}} \right].
\end{equation}
 \hrulefill
 \end{figure*}

\subsubsection{Spatially coupled PLDPC codes}
SC-PLDPC codes are constructed based on underlying PLDPC block codes (PLDPC-BCs).
We denote $W$ as the coupling width (equivalent to the aforementioned syndrome former memory $m_s$) and $L$ as the coupling length.
Based on the $m\times n$ protomatrix $\bm{B}$ of an underlying 
PLDPC-BC,
an edge-spreading procedure \cite{Mitchell2015} can be first used to obtain $W+1$ split protomatrices
$\bm{B}_i$ ($i=0,1,\ldots,W$) under the constraint $\bm{B} = \sum_{i = 0}^W {{\bm{B}_i}}$.
 Then  $L$ sets of such protomatrices are coupled to construct an
SC-PLDPC code \cite{Stinner2016}.
When the $L$ sets of protomatrices are coupled
and 
directly terminated, the resultant protomatrix equals 
\eqref{eq:td}. 
Such code is called {\color{black} an} SC-PLDPC terminated code (SC-PLDPC-TDC) and
its code rate equals
${R_{\rm SC-PLDPC-TDC}} = 1 - \frac{{L + W}}{L}\left( {1 - R_{\rm PLDPC-BC}} \right)$
where $R_{\rm PLDPC-BC} = 1 - \frac{m}{n}$ is the code rate of its underlying block code.
When the protograph of a spatially coupled code is terminated with ``end-to-end'' connections, the corresponding code is called SC-PLDPC tail-biting code (SC-PLDPC-TBC), whose protomatrix can be written as \eqref{eq:tbc}.
The code rate $R_{\rm SC-PLDPC-TBC}$ of {\color{black} an} SC-PLDPC-TBC is the same as that of its underlying block code, i.e.,
${R_{\rm SC-PLDPC-TBC}} = \frac{nL - mL}{nL} = R_{\rm PLDPC-BC}$.
By extending the coupling length $L$ of {\color{black} an} SC-PLDPC-TDC to infinity, {\color{black} an} 
SC-PLDPC-CC is formed.
The semi-infinite protomatrix of {\color{black} an} SC-PLDPC-CC is given by
\eqref{eq:cc_mat}.
The code rate of SC-PLDPC-CC equals that of the underlying block code  \cite{Stinner2016}, i.e.,
${R_{\rm SC-PLDPC-CC}} =
\lim_{L \to \infty} {R_{\rm SC-PLDPC-TDC}} = R_{\rm PLDPC-BC}$.
Once the protomatrix of a spatially coupled code is derived, a two-step lifting method 
can be used to construct a SC-PLDPC code.
{\color{black}(A single-step lifting method \cite{Battaglioni2021} can also be used to obtain an SC-PLDPC code.)}


\section{Spatially Coupled PLDPC-Hadamard Convolutional Codes}\label{sect:SC-PLDPCH-CC}
In this section, we show the details of our proposed 
SC-PLDPCH-CC. First, we show the way of constructing SC-PLDPCH codes, including SC-PLDPCH tail-biting code (SC-PLDPCH-TBC),
SC-PLDPCH terminated code (SC-PLDPCH-TDC) and SC-PLDPCH-CC, from its block code counterpart. Second, we briefly explain the encoding process of SC-PLDPCH-CCs.
Third,  we describe an efficient decoding algorithm for SC-PLDPCH-CC,
which combines the layered decoding used for decoding
PLDPCH-BC \cite{Zhang2021layer} and the pipeline decoding used for decoding SC-PLDPC-CC \cite{Costello2014}.
Fourth, we propose a layered PEXIT algorithm for evaluating the theoretical threshold
of {\color{black} an} SC-PLDPCH-TDC, which is then used to approximate the threshold of the corresponding SC-PLDPCH-CC.
Fifth, we propose a 
GA 
to optimize the SC-PLDPCH-TDC/SC-PLDPCH-CC designs
based on a given PLDPCH-BC.

\subsection{Code Construction}\label{sect:code_str}
SC-PLDPCH codes
are constructed in a similar way as SC-PLDPC codes.
We also denote the coupling width as $W$ and coupling length as $L$
in {\color{black} an} SC-PLDPCH code.
Given a 
PLDPCH-BC 
with a protomatrix $\bm B$,
we apply the edge-spreading procedure to split $\bm B$
 into $W+1$ protomatrices $\bm{B}_i$ ($i = 0,1,\ldots,W$) under the constraint $\bm{B} = \sum_{i = 0}^W {{\bm{B}_i}}$.
 Then we couple $L$ sets of these matrices to construct the protomatrix of a
 SC-PLDPCH code.
Similar to the SC-PLDPC codes described in Section \ref{sc-ldpc-cc},
{\color{black} an} SC-PLDPCH-TDC is formed if the coupled matrices are directly terminated;
{\color{black} an} SC-PLDPCH-TBC is formed if the coupled matrices are connected end-to-end;
and {\color{black} an} SC-PLDPCH-CC is formed if the coupling length $L$ becomes infinite.
Since the constructed protomatrices only represent the connections between P-VNs and H-CNs, 
SC-PLDPCH codes
have protomatrix structures similar to those of SC-PLDPC codes, i.e., \eqref{eq:td} for SC-PLDPCH-TDC;
\eqref{eq:tbc} for SC-PLDPCH-TBC; and
\eqref{eq:cc_mat} for SC-PLDPCH-CC.
Unlike the protographs of SC-PLDPC codes which consist of P-VNs and SPC-CNs,
the protographs of SC-PLDPCH codes {\color{black}contain} P-VNs and
 H-CNs connected with some appropriate D1H-VNs.
Fig.~\ref{fig:PLDPCH_1}(b)
shows the protograph of {\color{black} an} SC-PLDPCH-CC derived from
the PLDPCH-BC in Fig.~\ref{fig:PLDPCH_1}(a).
%
%
 SC-PLDPCH codes can be constructed from the coupled protographs using the lifting process.
 Assuming that $\bm{B}$ has a constant row weight of $d$ and hence
an order-$r$ ($=d-2$) Hadamard code is used (recall $r$ is even), it can be readily shown that
the code rates of the SC-PLDPCH codes are as follows.
For SC-PLDPCH-TDCs the code rate equals
$R^{\rm even}_{\rm SC-PLDPCH-TDC} = \frac{{nL - m\left( {L + W} \right)}}{{nL + m\left( {L + W} \right)\left( {{2^r} - d} \right)}} = \frac{{n - m\left( {1 + \frac{W}{L}} \right)}}{{n + m\left( {1 + \frac{W}{L}} \right)\left( {{2^r} - d} \right)}}$.
For SC-PLDPCH-TBCs and SC-PLDPCH-CCs, their code rates are the same
as the block code counterparts, i.e.,
$ R^{\rm even}_{\rm SC-PLDPCH-TBC}  = R^{\rm even}_{\rm SC-PLDPCH-CC}
 =  R^{\rm even}_{\rm PLDPCH-BC} = \frac{{n - m}}{{n + m\left( {{2^r} - r - 2} \right)}}$.

\begin{figure*}[t]
\centerline{
\includegraphics[width=0.45\columnwidth]{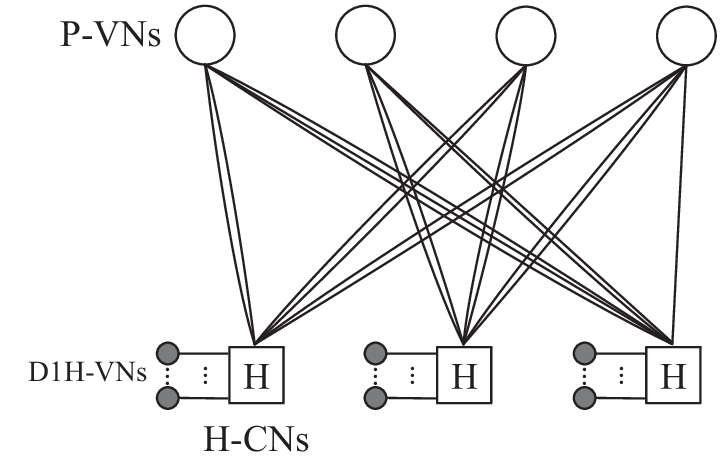}\;\;\;\;\;\;
\includegraphics[width=0.5\columnwidth]{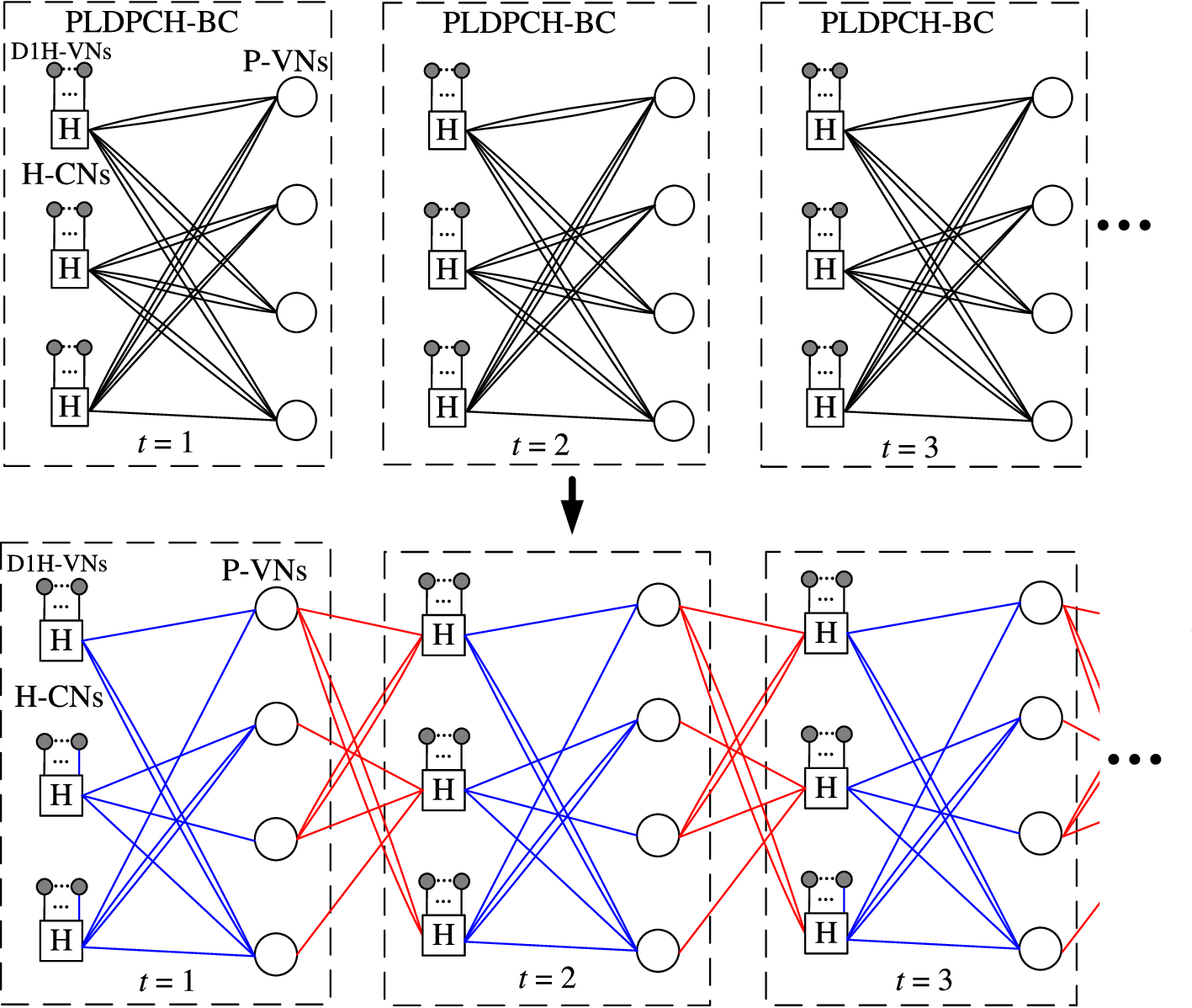}} 
 \centerline{(a) \rule{4cm}{0cm} (b)}
 \caption{(a) A protograph of PLDPC-Hadamard code. Number of D1H-VNs connected to each HCN is $2^r - d = 10$ using order-$r = d- 2=4$ Hadamard code. 
 (b) Protograph of {\color{black} an} SC-PLDPCH-CC derived from
the PLDPC-Hadamard code.}
  \label{fig:PLDPCH_1}
\end{figure*}

\subsection{Encoding of SC-PLDPCH-CC}
From this point forward and unless otherwise stated, we focus our study on SC-PLDPCH-CC.
We also assume that the row weight of
$\bm{B}$ equals $d=r+2$ and is even.
After performing a two-step lifting process on \eqref{eq:cc_mat},
we obtain the semi-infinite parity-check matrix of {\color{black} an} SC-PLDPCH-CC in Fig. \ref{fig:PLDPCH_CC_enc}.
Denoting the two lifting factors by  $z_1$ and $z_2$,
each $\bm{H}_i$ ($i=0,1,\ldots,W$) which is the lifted $\bm{B}_i$ has a size of $M \times N = m z_1 z_2 \times n z_1 z_2$.
At time $t$, $M-N$ information bits denoted by $\bm{b}(t) \in \{0,1\}^{M-N}$ are input to the SC-PLDPCH-CC encoder.
The output of the SC-PLDPCH-CC encoder contains  $N$ coded bits corresponding to P-VNs, which are denoted by $\bm{P}(t)$;
 and $M(2^r - r - 2)$ Hadamard parity-check bits corresponding to D1H-VNs, which are denoted by $\bm{D}(t)$.
Referring to Fig. \ref{fig:PLDPCH_CC_enc}, we generate the output bits as follows.
\begin{enumerate}
\item $t=1$: Given $\bm{b}(1)$, $\bm{P}(1)$ is generated based on the first block row of ${\bm{H}_{\rm SC-PLDPCH-CC}}$, i.e., $\bm{H}_0$.
Moreover, $\bm{D}(1)$  is computed based on $[ {\overbrace {\bm{0}\ \cdots \bm{0}}^{_W}\ \bm{P}(1)} ]$ and the structure $[{\bm{H}_W} \ \cdots \ {\bm{H}_1} \ {\bm{H}_0}]$, where $\bm{0}$ is the length-$N$ zero vector. 

\item $t=2$: Given $\bm{b}(2)$ and $\bm{P}(1)$, $\bm{P}(2)$ is generated based on the second block row of ${\bm{H}_{\rm SC-PLDPCH-CC}}$, i.e., $[\bm{H}_1 \ \bm{H}_0]$. Moreover, $\bm{D}(2)$  is computed based on $[ {\overbrace {\bm{0}\ \cdots \bm{0}}^{_{W-1}}\ \bm{P}(1)\ \bm{P}(2)} ]$ and the structure $[{\bm{H}_W} \ \cdots \ \bm{H}_1 \ {\bm{H}_0}]$.

 \item$t \le W$: Given $\bm{b}(t)$ and $[\bm{P}(1) \ \bm{P}(2) \ \cdots \ \bm{P}(t-1)]$,
$N$ coded bits $\bm{P}(t)$ are generated based on the $t$-th block row of ${\bm{H}_{\rm SC-PLDPCH-CC}}$,
i.e., $[\bm{H}_{t-1} \  \cdots  \ {\bm{H}_1} \ \bm{H}_0]$.
$\bm{D}(t)$  corresponding to the $M(2^r-r-2)$ D1H-VNs are computed based on $[\overbrace {\bm{0}\ \cdots \bm{0}}^{_{W+1-t}} \bm{P}(1) \ \cdots \ \bm{P}(t)]$ and the structure $[{\bm{H}_W} \ \cdots \ {\bm{H}_1} \ {\bm{H}_0}]$.

\item$t > W$: Given $\bm{b}(t)$ and $[\bm{P}(t-W) \ \bm{P}(t-W+1) \ \cdots \ \bm{P}(t-1)]$,
$\bm{P}(t)$
is generated based on the $t$-th block row of ${\bm{H}_{\rm SC-PLDPCH-CC}}$,
i.e., $[\bm{H}_{W} \  \cdots  \ {\bm{H}_1} \ \bm{H}_0]$.
Then, $\bm{D}(t)$ is computed based on $[\bm{P}(t-W) \ \cdots \  \bm{P}(t-1) \ \bm{P}(t)]$ and the structure $[{\bm{H}_W} \ \cdots \ {\bm{H}_1} \ {\bm{H}_0}]$.
{\color{black} The constraint length of the SC-PLDPCH-CC therefore equals 
$(W+1)N+M(2^r - r - 2)$.}
\end{enumerate}

\noindent \textit{Remarks:} The values of $\bm{D}(t)$ are generated during the encoding corresponding to the $t$-th block row. They are not needed for generating
other $\bm{D}(t')$ where $t\ne t'$.
When $t \le W$, $W+1-t$ length-$N$ zero vectors are inserted in front of $\bm{P}(1)$ for computing $\bm{D}(t)$. But these zero vectors are not transmitted through the channel.

\begin{figure}[t]
\centerline{
\includegraphics[width=0.6\columnwidth]{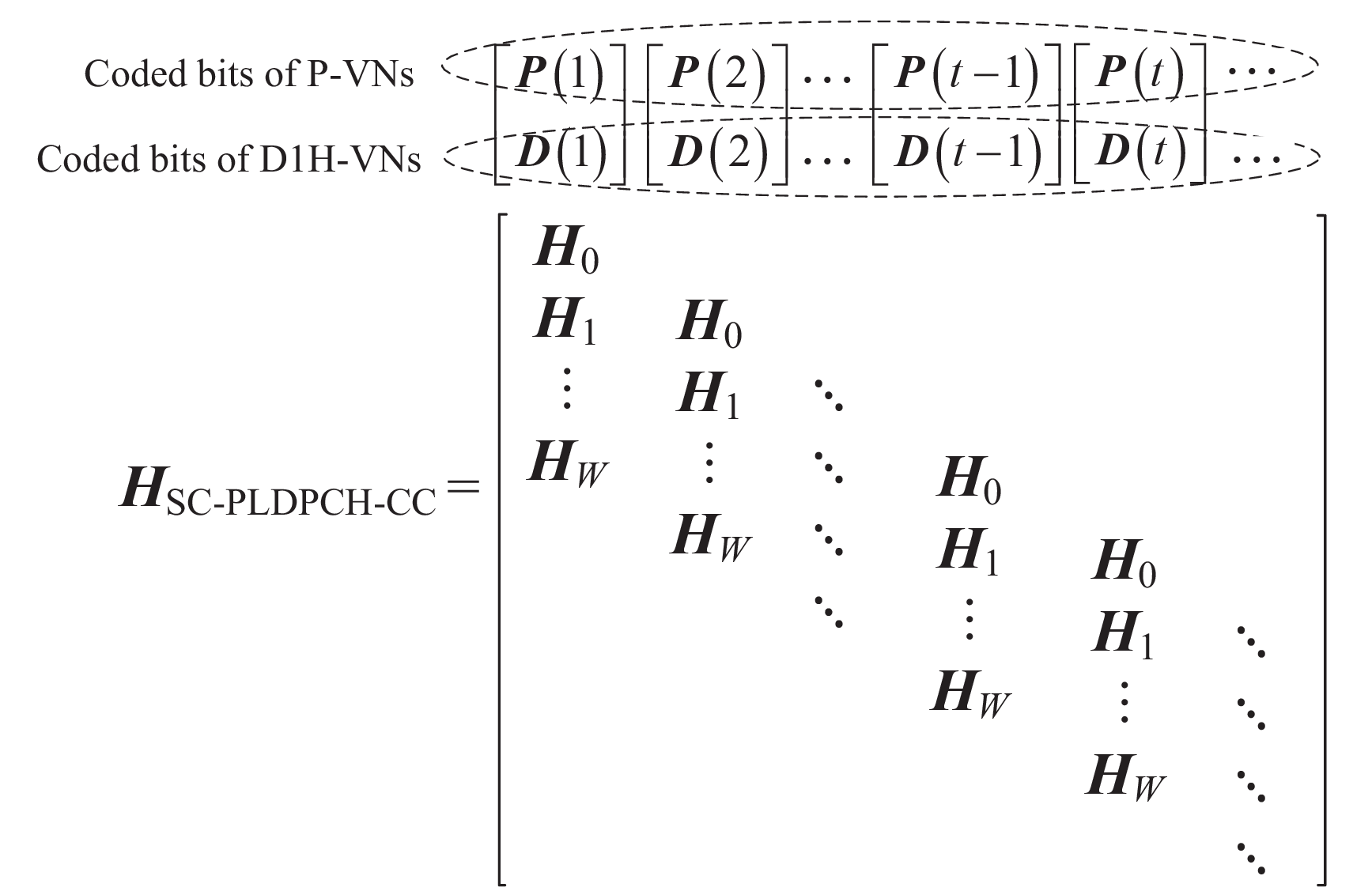}}
\caption{Encoding of {\color{black} an} SC-PLDPCH-CC. Coded bits $\bm{P}(1)$, $\bm{P}(2)$, $\ldots$, $\bm{P}(t-1)$, $\bm{P}(t)$, $\ldots$ correspond to P-VNs at time $1,2,\ldots,t-1,t,\ldots$. Hadamard parity-check bits $\bm{D}(1), \bm{D}(2), \ldots, \bm{D}(t-1), \bm{D}(t), \ldots$ correspond to D1H-VNs at time $1,2,\ldots,t-1,t,\ldots$.}
  \label{fig:PLDPCH_CC_enc}
\end{figure}

\begin{figure*}[t]
\centerline{
\includegraphics[width=1\columnwidth]{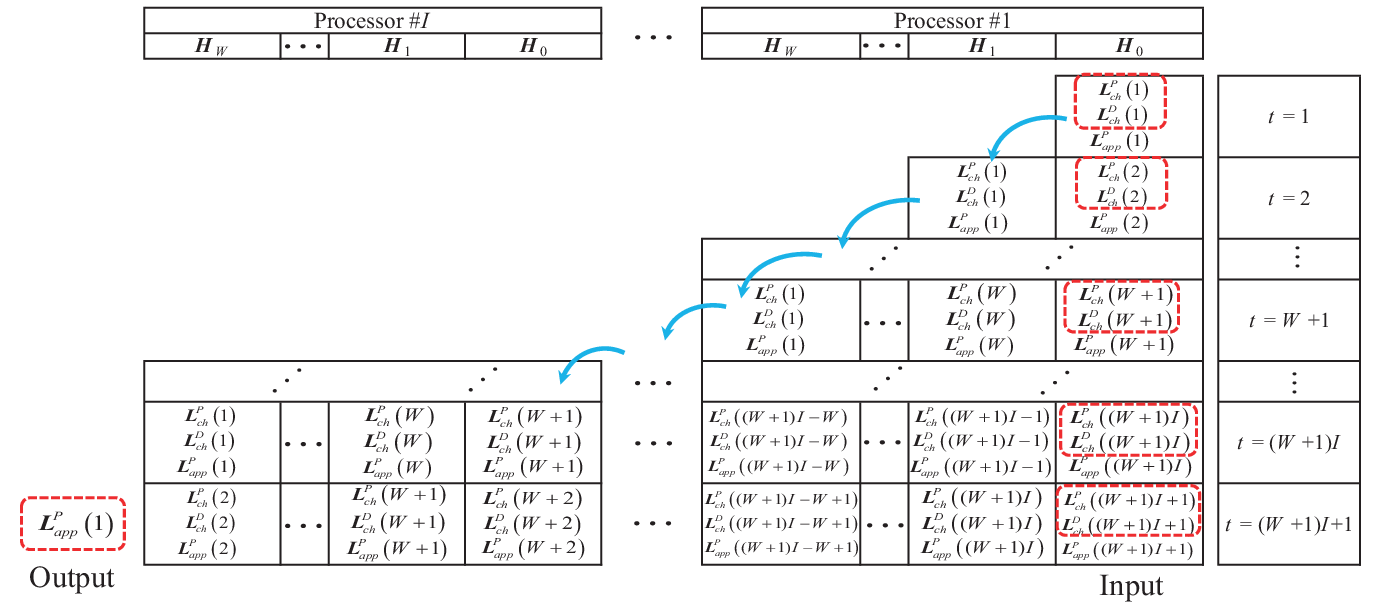}}
\caption{Structure of a pipeline SC-PLDPCH-CC decoder consisting of $I$ processors (PLDPC-Hadamard block sub-decoders).
$\{ \bm{L}_{ch}^P(t), \bm{L}_{ch}^D(t)\}$ $(t=1,2,\ldots)$ are input into the decoder one set by one set. Every time, all sets of LLRs inside the decoder are
shifted to the left, and all APP-LLRs of all P-VNs inside the different $I$ processors
 are updated. When $\{\bm{L}_{ch}^P((W+1)I+t'),\bm{L}_{ch}^D((W+1)I+t')\}$ $(t'=1,2,\ldots)$ is input to the decoder,  the APP-LLRs $\bm{L}_{app}^P(t')$  are output and the
values of the coded bits $\bm{P}(t')$ are determined. }
  \label{fig:PLDPCH_CC_dec}
\end{figure*}

\subsection{Pipeline Decoding}\label{pipeline_dec}

\subsubsection{Decoding Structure}
At the receiving end, we receive channel observations regarding the coded bits $\bm{P}(t)$ (corresponding to P-VNs) and Hadamard parity-check bits $\bm{D}(t)$  (corresponding to D1H-VNs).
We denote the log-likelihood-ratio (LLR) values
corresponding to $\bm{P}(t)$ by $\bm{L}_{ch}^P(t)$ and the LLR values
corresponding to $\bm{D}(t)$ by $\bm{L}_{ch}^D(t)$.
We consider a pipeline decoder which consists of $I$ identical message-passing processors \cite{Jimenez1999,Sham2013,Costello2014}.
Each processor is a PLDPC-Hadamard block sub-decoder corresponding to $[{\bm{H}_W} \ \cdots \ {\bm{H}_1} \ {\bm{H}_0}]$.
Thus, each processor operates on $W+1$ sets of P-VNs and
one set of D1H-VNs each time, i.e., a total of $N(W+1)$ P-VNs
and $M(2^r-r-2)$ D1H-VNs (when $r$ is even).
Hence the pipeline decoder operates on $(W+1)I$ sets of P-VNs and $I$ sets of D1H-VNs each time.
Each processor (sub-decoder) can apply either the standard decoding algorithm or the layered decoding algorithm to compute/update the \emph{a posteriori} probability in LLR form (APP-LLR) for the coded bits $\bm{P}(t)$ and the related extrinsic LLR information.
Here, we apply the layered decoding algorithm \cite{Zhang2021layer} in each of these
PLDPC-Hadamard block sub-decoders.

We denote the
APP-LLR values of the coded bits $\bm{P}(t)$ by
$\bm{L}_{app}^P(t)$.
Referring to Fig.~\ref{fig:PLDPCH_CC_dec},  $\{\bm{L}_{ch}^P(1),\bm{L}_{ch}^D(1)\}$
is first input to the pipeline decoder and Processor \#1 updates the
APP-LLR of all P-VNs inside, i.e., $\bm{L}_{app}^P(1)$.
Also, extrinsic LLR information is updated and stored in the processor but is not
depicted in the figure.
Then,  $\{\bm{L}_{ch}^P(2),\bm{L}_{ch}^D(2)\}$
is input to the pipeline decoder while
 $\{\bm{L}_{ch}^P(1),\bm{L}_{ch}^D(1),\bm{L}_{app}^P(1)\}$ and related  extrinsic LLR information are shifted to the left in the decoder.
Processor \#1 updates the
APP-LLRs of all P-VNs inside, i.e., $\bm{L}_{app}^P(1)$ and $\bm{L}_{app}^P(2)$.
Again, extrinsic LLR information is updated and stored in the processor but is not
depicted. Subsequently, $\{ \bm{L}_{ch}^P(t), \bm{L}_{ch}^D(t)\}$ $(t=3,4,\ldots)$ are input into the decoder one set by one set. Every time, all sets of LLRs inside the decoder are
shifted to the left by one ``$\bm{H}_i$'' block, and all APP-LLRs of all P-VNs inside the different $I$ processors
 are updated.
When
$\{\bm{L}_{ch}^P((W+1)I+1),\bm{L}_{ch}^D((W+1)I+1)\}$
 is input to the pipeline decoder, the APP-LLRs $\bm{L}_{app}^P(1)$
have gone through the iterative process and are output from the decoder. Hard decisions are made based on these APP-LLRs to determine the values of the coded bits $\bm{P}(1)$.
The process continues and every time $\{\bm{L}_{ch}^P((W+1)I+t'),\bm{L}_{ch}^D((W+1)I+t')\}$ $(t'=1,2,\ldots)$ is input to the decoder,  the APP-LLRs $\bm{L}_{app}^P(t')$  are output and the
values of the coded bits $\bm{P}(t')$ are determined.
{\color{black}
\subsubsection{Latency and Complexity Analysis}\label{sect:CA}
In this section, we compare the latency and complexity of
 SC-PLDPCH-CC decoder and PLDPCH-BC decoder  \cite{Zhang2021layer,zhang2021HWPLDPCH}.
{\color{black} We define ``latency'' as the time interval between
(i) the LLR information entering a decoder
 and (ii) the corresponding decoded P-VNs output from the decoder. }
  We assume a two-step lifting process with the factors $z_1$ and $z_2$
 applied to lift the base matrices of SC-PLDPCH-CC and PLDPCH-BC.
Moreover, layered decoding is used in both cases.
For a given PLDPCH-BC, we denote the time taken to complete one
 decoding iteration by $T_r$, which has been shown
 to be proportional to the number of layers, i.e., $m z_1$  \cite{zhang2021HWPLDPCH}.
 Denoting  the
maximum number of iterations by $I_{BC}$,
the 
{\color{black} time delay (i.e., latency)} for decoding one PLDPCH-BC codeword equals $T_{\rm PLDPCH-BC}= I_{BC} T_r $.

Based on the same protomatrix as the PLDPCH-BC, we use the edge spreading approach to construct an SC-PLDPCH-CC.
We exploit the pipeline decoding method in the previous section with $I$ identical processors, each of which applies the layered decoding algorithm.
Moreover, it can be readily shown that the design used to implement the
aforementioned PLDPCH-BC decoder \cite{zhang2021HWPLDPCH}   can be slightly modified
 to implement these processors.
As each processor needs to update $m z_1$ layers of information
whenever a new set of $\{ \bm{L}_{ch}^P(t), \bm{L}_{ch}^D(t)\}$ $(t=1,2,\ldots)$ enters the decoder,
the time taken is the same as for the PLDPCH-BC
 decoder to complete one iteration, i.e., $T_r$.
As shown in  Fig. \ref{fig:PLDPCH_CC_dec} and discussed in
the previous section, the APP-LLRs $\bm{L}_{app}^P(t')$
 are output from the decoder when
$\{\bm{L}_{ch}^P((W+1)I+t'),\bm{L}_{ch}^D((W+1)I+t')\}$
 is input to the pipeline decoder.
Thus, the 
{\color{black} total time delay (i.e., latency)} for decoding one set of input equals $T_{\rm SC-PLDPCH-CC}=(W+1) I T_r$.
In addition, $(W+1)I$ sets of LLRs corresponding to different $t$'s are being processed
by the pipeline decoder, as compared to only one set of LLRs being processed by
the PLDPCH-BC decoder at any time.

In summary, when a PLDPCH-BC and an SC-PLDPCH-CC are ``derived'' from the
same protomatrix of size $m \times n$ with the same lifting factors $z_1$ and $z_2$, the  PLDPCH-BC decoder
and each processor in an SC-PLDPCH-CC pipeline decoder processes the
$m z_1$ layers in layered decoding with the same time delay.
The SC-PLDPCH-CC pipeline decoder consists of $I$ processors, each having a similar structure
as a PLDPCH-BC decoder; and requires $(W+1)I$ times memory storage compared with
a PLDPCH-BC decoder.
The latencies of the SC-PLDPCH-CC pipeline decoder and PLDPCH-BC decoder
are $T_{\rm SC-PLDPCH-CC}=(W+1) I T_r$ and $T_{\rm PLDPCH-BC}= I_{BC} T_r$, respectively.
Under the condition that the two latencies are identical, i.e.,
$T_{\rm SC-PLDPCH-CC}=T_{\rm PLDPCH-BC}$ or $(W+1) I = I_{BC}$,
the SC-PLDPCH-CC pipeline decoder achieves a $(W+1) I$ times throughput
compared with the PLDPCH-BC decoder.
}
{\color{black}
\subsubsection{SC-PLDPCH-CC and PLDPCH-BC under the same constraint length/blocklength}\label{sect:same_length}
In this section, we further compare the case when the constraint length of a
 SC-PLDPCH-CC  is the same as the blocklength of a PLDPCH-BC.
 We assume a two-step lifting process with the factors $z_1$ and $z_2$ applied to lift the $m\times n$ base matrices of SC-PLDPCH-CC and, $z_1'$ and $z_2'$ applied to lift the same size base matrix of PLDPCH-BC.
The constraint length of the SC-PLDPCH-CC is
 $L_{\rm CC-CL} = (W+1)N+M(2^r - r - 2)=(W+1)z_1 z_2 n + z_1 z_2 m(2^r-r-2)$
 and the  blocklength of the PLDPCH-BC is 
  $L_{\rm BC-BL}=z_1' z_2' n + z_1' z_2' m(2^r-r-2)$. 
For simplicity, we let $z_1 = z_1'$. When $L_{\rm CC-CL} =L_{\rm BC-BL}$,
 we have $\delta_{z_2} \triangleq z_2' / z_2 = 1+ \frac{W}{1 + (2^r-r-2)\frac{m}{n}}$\footnote{\color{black} Note that when $r$ is odd,  $\delta_{z_2}  = 1+ \frac{W}{1 + (2^r-2)\frac{m}{n}}$}. 
 The result implies that
 $z_2'$ is strictly larger than $z_2$.
For example, when $W=1, m=7, n=11, r=4$, we obtain 
 $\delta_{z_2} = z_2'/z_2 = 1.136$.
When the number of H-CNs in one layer 
is increased by a factor of $\delta_{z_2}$,
the time taken to complete one
 decoding iteration is increased by the same factor.
In order to maintain the same decoding latency,  
  the maximum number of iterations 
  for decoding one PLDPCH-BC codeword should be reduced by the same factor, i.e.,
reduced from $I_{BC}$ (see above section) to  $I'_{BC} = I_{BC} / \delta_{z_2}$.
}

\begin{figure*}[!t]
\begin{equation}\label{mat:th_r4_b0}{
{\bm{B}_0} = \left[ {\begin{array}{*{20}{c}}
1&    0&    0&    0&    0&    0&    0&    0&    0&    0&    1\\
0&    0&    1&    0&    0&    0&    0&    0&    0&    1&    0\\
1&    0&    0&    0&    0&    0&    0&    0&    0&    0&    0\\
0&    0&    0&    3&    0&    0&    0&    0&    0&    2&    0\\
0&    0&    0&    0&    0&    0&    0&    0&    0&    0&    0\\
1&    0&    0&    2&    0&    0&    1&    0&    0&    0&    0\\
0&    0&    0&    0&    0&    0&    0&    0&    1&    0&    0
\end{array}} \right];
\;\;\;\;\;
{\bm{B}_1} = \left[ {\begin{array}{*{20}{c}}
0&    0&    0&    0&    0&    0&    1&    0&    3&    0&    0\\
0&    1&    1&    0&    0&    0&    0&    0&    0&    1&    1\\
1&    1&    0&    0&    1&    1&    0&    0&    0&    0&    1\\
0&    1&    0&    0&    0&    0&    0&    0&    0&    0&    0\\
2&    0&    0&    0&    0&    0&    0&    1&    0&    3&    0\\
2&    0&    0&    0&    0&    0&    0&    0&    0&    0&    0\\
1&    0&    0&    1&    1&    0&    0&    0&    0&    2&    0
\end{array}} \right]}
\end{equation}
\begin{equation}\label{mat:th_r4_b0_2}{
{\bm{B}_0} = \left[ {\begin{array}{*{20}{c}}
0&    0&    0&    0&    0&    0&    0&    0&    0&    0&    0\\
0&    1&    2&    0&    0&    0&    0&    0&    0&    0&    0\\
2&    1&    0&    0&    1&    0&    0&    0&    0&    0&    1\\
0&    0&    0&    3&    0&    0&    0&    0&    0&    1&    0\\
0&    0&    0&    0&    0&    0&    0&    1&    0&    3&    0\\
2&    0&    0&    1&    0&    0&    0&    0&    0&    0&    0\\
1&    0&    0&    1&    0&    0&    0&    0&    0&    0&    0
\end{array}} \right];\;\;\;
{\bm{B}_1} = \left[ {\begin{array}{*{20}{c}}
1&    0&    0&    0&    0&    0&    1&    0&    3&    0&    1\\
0&    0&    0&    0&    0&    0&    0&    0&    0&    2&    1\\
0&    0&    0&    0&    0&    1&    0&    0&    0&    0&    0\\
0&    1&    0&    0&    0&    0&    0&    0&    0&    1&    0\\
2&    0&    0&    0&    0&    0&    0&    0&    0&    0&    0\\
1&    0&    0&    1&    0&    0&    1&    0&    0&    0&    0\\
0&    0&    0&    0&    1&    0&    0&    0&    1&    2&    0
\end{array}} \right]}
\end{equation}
 \hrulefill
\end{figure*}

\subsection{Layered PEXIT Algorithm}\label{sect:layered_PEXIT}
In \cite{zhang2021}, a low-complexity PEXIT chart technique has been proposed to evaluate the theoretical threshold of PLDPCH-BCs.
The thresholds of SC-PLDPCH-CCs are expected to be comparable to those of their underlying block codes.
However, direct analysis of SC-PLDPCH-CCs with infinite length is very complicated and time-consuming, which is not conducive to the optimal design of the codes in
 the next section, i.e., Section \ref{sect:GA}.

As mentioned in Section \ref{sect:code_str}, under the same set of split protomatrices $\{\bm{B}_0, \bm{B}_1, \ldots, \bm{B}_W\}$, SC-PLDPCH-CCs can be obtained by extending the 
coupling length 
$L$ 
of SC-PLDPCH-TDCs to infinity.
{\color{black} An} SC-PLDPCH-TDC can be treated as a PLDPCH-BC with a large size and hence its threshold
 can be evaluated using the original PEXIT chart method in \cite{zhang2021}.
For SC-PLDPCH-TDCs constructed with different split protomatrices, the original PEXIT chart method will generate different (i.e., distinguishable) thresholds \footnote{For SC-PLDPCH-TDCs constructed with different split protomatrices but arrived at
the same threshold, we further verify their error performance by simulations.}.
For example, Fig. \ref{fig:th_r4} shows the thresholds for two different SC-PLDPCH-TDCs using the original PEXIT chart method.
We set $W = 1$. Based on the optimal protomatrix $\bm{B}$ in \cite{zhang2021},
the split protomatrices of SC-PLDPCH-TDC \#1 are
shown as \eqref{mat:th_r4_b0} 
and
the split protomatrices of SC-PLDPCH-TDC \#2 are
shown as \eqref{mat:th_r4_b0_2},  
where $\bm{B}_0 + \bm{B}_1 = \bm{B}$.
From Fig. \ref{fig:th_r4}, we can observe that the thresholds of two SC-PLDPCH-TDCs are distinguishable and
{\color{black} improved
{(reduced)} as the coupling length $L$ increases.
Moreover, the improvement diminishes as  $L$ becomes large.
Thus, threshold saturation is observed. }
Subsequently, we select the protomatrices with good thresholds (i.e., low $E_b/N_0$) to construct the corresponding SC-PLDPCH-CCs.

\begin{figure}[t]
\centerline{
\includegraphics[width=0.7\columnwidth]{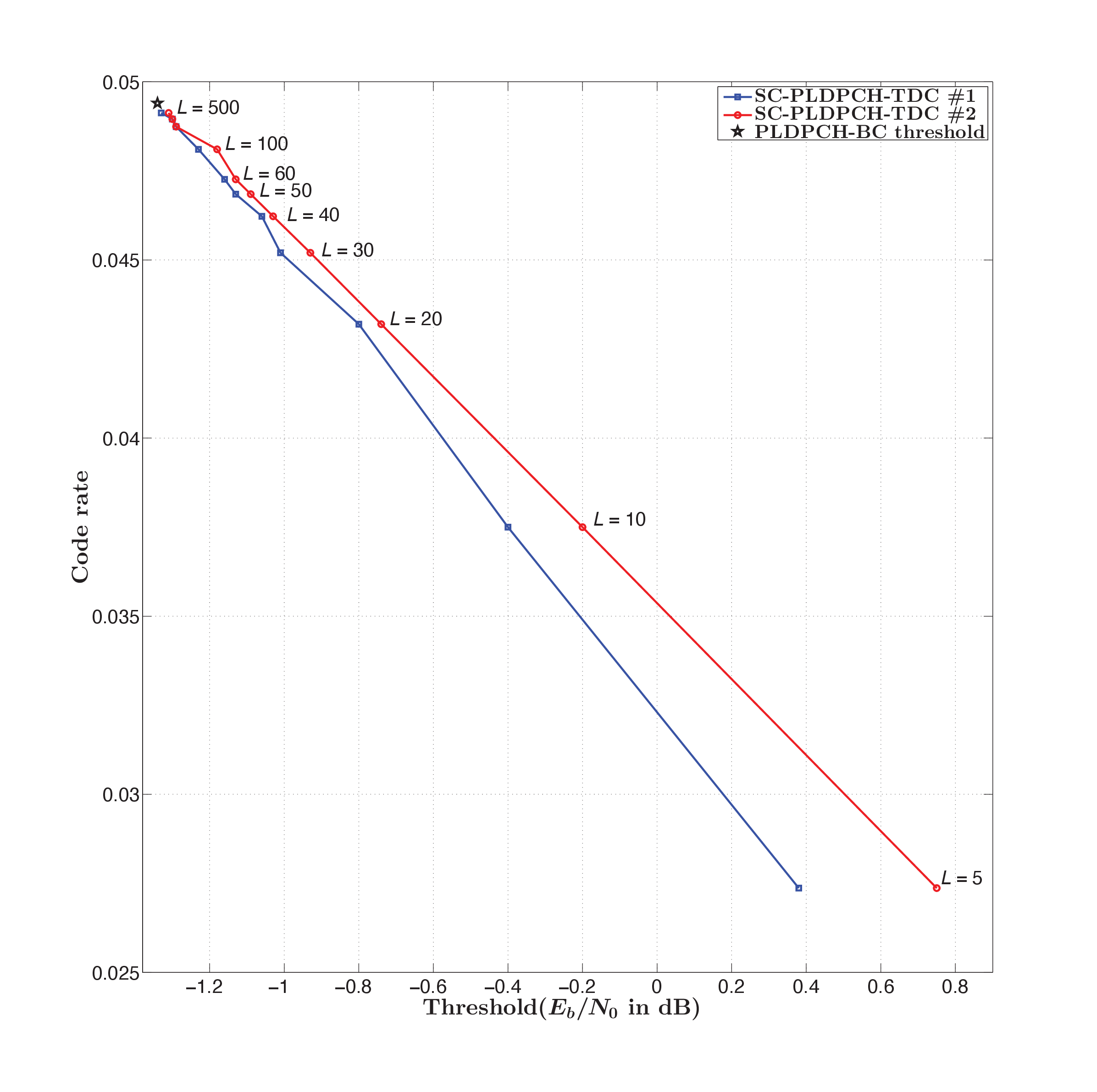}}
\caption{Distinguishable thresholds for two different SC-PLDPCH-TDCs corresponding to different set of protomatrices $\{\bm{B}_0, \bm{B}_1\}$. $W = 1$. \color{black}Their underlying block code is the optimal $r = 4$ PLDPCH-BC  \cite{zhang2021} with the decoding threshold indicated by the symbol $\star$. }
  \label{fig:th_r4}
\end{figure}

To improve the convergence rate of the original PEXIT chart method,
we propose a layered PEXIT method to analyze the SC-PLDPCH-TDCs.
We first define 
 $I_{av}(i,j)$ as the \textit{a priori} mutual information (MI)  from the $i$-th H-CN to the  $j$-th P-VN;
$I_{ev}(i,j)$ as the extrinsic MI from the $j$-th P-VN to the $i$-th H-CN;
 $I_{ah}(k)$ as the \textit{a priori} MI of the $k$-th information bit in the $i$-th H-CN;
 $I_{eh}(k)$ as the extrinsic MI of the $k$-th information bit in the $i$-th H-CN;
 $I_{app}(j)$ as the \textit{a posteriori} MI of the $j$-th P-VN;
 $\sigma_{app}(j)$ as the \emph{a posteriori} information of the $j$-th P-VN; and
 $\sigma_{temp}(j)$ as the temporary value of $\sigma_{app}(j)$.
 We also assume that the channel LLR value $L_{ch}$ follows a normal distribution  $\mathcal{N}(\sigma_{L_{ch}}^{2}/2,\sigma_{L_{ch}}^{2})$, where $\sigma_{L_{ch}}^2 = 8R \cdot E_b/N_0$, $R$ is the code rate of SC-PLDPCH-TDC, and $E_b/N_0$ is the bit-energy-to-noise-power-spectral-density ratio.
When the output MI from P-VN or H-CN processors is  $I_{\rm MI}$, we suppose that the corresponding
LLR values of the extrinsic information obey a normal distribution $\mathcal{N}(\pm \sigma^2/2, \sigma^2)$.
The relationship between $I_{\rm MI}$ and $\sigma$ can be approximately computed by
the functions $I_{\rm MI} = J(\sigma)$ and $\sigma = J^{-1}(I_{\rm MI})$ in \cite{Fang2015,Brink2001}.

 \begin{figure*}[t]
 \begin{eqnarray}
\label{eq:example_td}
{\bm{B}} &=& \left[ {\begin{array}{*{20}{c}}
2&0&2&2\\
0&2&2&2\\
3&2&0&1
\end{array}} \right],\
{\bm{B}_0} = \left[ {\begin{array}{*{20}{c}}
1&0&0&2\\
0&1&1&1\\
1&2&0&1
\end{array}} \right],\
{\bm{B}_1} = \left[ {\begin{array}{*{20}{c}}
1&0&2&0\\
0&1&1&1\\
2&0&0&0
\end{array}} \right]\\
 {\bm{B}_{\rm SC-PLDPCH-TDC}} &=& \left[ {\begin{array}{*{20}{c}}
{{\bm{B}_0}}&{\bm{0}}\\
{{\bm{B}_1}}&{{\bm{B}_0}}\\
{\bm{0}}&{{\bm{B}_1}}
\end{array}} \right] = \left[ {\begin{array}{*{20}{c}}
\cellcolor{green}1&\cellcolor{green}0&\cellcolor{green}0&\cellcolor{green}2&{0}&{0}&{0}&{0}\\
\cellcolor{green}0&\cellcolor{green}1&\cellcolor{green}1&\cellcolor{green}1&{0}&{0}&{0}&{0}\\
\cellcolor{green}1&\cellcolor{green}2&\cellcolor{green}0&\cellcolor{green}1&{0}&{0}&{0}&{0}\\
\cellcolor{yellow}1&\cellcolor{yellow}0&\cellcolor{yellow}2&0\cellcolor{yellow}&\cellcolor{green}1&\cellcolor{green}0&\cellcolor{green}0&\cellcolor{green}2\\
\cellcolor{yellow}0&\cellcolor{yellow}1&\cellcolor{yellow}1&\cellcolor{yellow}1&\cellcolor{green}0&\cellcolor{green}1&\cellcolor{green}1&\cellcolor{green}1\\
\cellcolor{yellow}2&\cellcolor{yellow}0&\cellcolor{yellow}0&\cellcolor{yellow}0&\cellcolor{green}1&\cellcolor{green}2&\cellcolor{green}0&\cellcolor{green}1\\
{0}&{0}&{0}&{0}&\cellcolor{yellow}1&\cellcolor{yellow}0&\cellcolor{yellow}2&\cellcolor{yellow}0\\
{0}&{0}&{0}&{0}&\cellcolor{yellow}0&\cellcolor{yellow}1&\cellcolor{yellow}1&\cellcolor{yellow}1\\
{0}&{0}&{0}&{0}&\cellcolor{yellow}2&\cellcolor{yellow}0&\cellcolor{yellow}0&\cellcolor{yellow}0
\end{array}} \right]
\label{eq:example_B_TDC} \\
 {\sigma _{temp}}(j) &=&  {\sqrt {({\sigma _{app}}\left( j \right))^2 - b_{\rm TDC}(i,j) \times {{\left( {{J^{ - 1}}\left( {{I_{av}}\left( {i,j} \right)} \right)} \right)}^2}} }  \ \forall \ j \label{eq:sigma_app_1} \\
 I_{ev}\left( {i,j} \right) &=& J\left( {\sqrt {({\sigma _{app}}\left( j \right))^2 - {{\left( {{J^{ - 1}}\left( {{I_{av}}\left( {i,j} \right)} \right)} \right)}^2}} } \right) \ \forall \ j \label{eq:I_ev} \\
\bm{b}_{\rm TDC}(1,:) &=&  \left[ {\begin{array}{*{20}{c}}
1&0&0&2&0&0&0&0
\end{array}} \right] \label{eq:example_1} \\
\bm{I}_{ev}(1,:) &=& \left[ {\begin{array}{*{20}{c}}
{{I_{ev}}\left( {1, 1} \right)}&0&0&{{I_{ev}}\left( {1,4} \right)}&0&0&0&0\\
\end{array}} \right]\label{eq:I_ev_ex_1} \\
\bm{I}_{ah} &=& \left[ {\begin{array}{*{20}{c}}
{\color{black}{I_{ah}}\left( {1} \right)}& {\color{black}{I_{ah}}\left( {2} \right)}&{\color{black}{I_{ah}}\left( {3} \right)}&{{I_{ah}}\left( {4} \right)}&{{I_{ah}}\left( {5} \right)}& {{I_{ah}}\left( {6} \right)}\\
\end{array}} \right] \nonumber\\
&=&\left[ {\begin{array}{*{20}{c}}
{\color{black}1}&{\color{black}1}&{\color{black}1}&{{I_{ev}}\left( {1,1} \right)}&{{I_{ev}}\left( {1,4} \right)}&{{I_{ev}}\left( {1,4} \right)}\\
\end{array}} \right]  \label{eq:I_ah_1} \\
{I_{eh}}\left( {k} \right)
 &=& \frac{1}{2}\sum\limits_{x \in \{ 0, 1\} } {\int_{ - \infty }^\infty  {{p_e}\left( {\xi |X = x} \right)  } }
{{{\log }_2}\frac{{2 \cdot {p_e}\left( {\xi |X = x} \right)}}{{{p_e}\left( {\xi |X =  ``0"} \right) + {p_e}\left( {\xi |X = ``1"} \right)}}}d\xi \label{I_eh_k_old}\\
 &\approx& 1 - \frac{1}{w}\sum\limits_{\alpha  = 1}^w {{{\log }_2}\left( {1 + {e^{ - \left( {1 - 2U\left( {\alpha ,k} \right)} \right) \times V\left( {\alpha ,k} \right)}}} \right)} \label{I_eh_k}\\
\bm{I}_{eh}
&=& \left[ {\begin{array}{*{20}{c}}
{\color{black}{I_{eh}}\left( {1} \right)}&{\color{black}{I_{eh}}\left( {2} \right)}&{\color{black}{I_{eh}}\left( {3} \right)}&{{I_{eh}}\left( {4} \right)} &{{I_{eh}}\left( {5} \right)}&{{I_{eh}}\left( {6} \right)}
\end{array}} \right] \label{I_eh_1}  \\
\bm{I}_{av}(1,:) &=& \left[ {\begin{array}{*{20}{c}}
{{I_{av}}\left( {1,1} \right)}&0&{{I_{av}}\left( {1,3} \right)}&0&0&0&0&0&\\
\end{array}} \right] \nonumber\\
&=&
 \left[ {\begin{array}{*{20}{c}}
I_{eh}(4)&0&\frac{1}{2}\sum\limits_{k = 5}^6 I_{eh}(k)&0&0&0&0&0
\end{array}} \right]
\label{eq:I_av_1}\\
{\sigma _{app}}\left( j \right) &=&  {\sqrt {({\sigma _{temp}}\left( j \right))^2 + b_{\rm TDC}(i,j) \times {{\left( {{J^{ - 1}}\left( {{I_{av}}\left( {i,j} \right)} \right)} \right)}^2}} }  \ \forall \ j \label{eq:sigma_app_2}
\end{eqnarray}
 \hrulefill
 \end{figure*}

Given a coupling width $W$ and a set of protomatrices $\{\bm{B}_0, \bm{B}_1, \ldots, \bm{B}_W\}$ each of size $m \times n$, we use these protomatrices to construct {\color{black} an} SC-PLDPCH-TDC with the coupling width $W$ and an appropriate coupling length $L$.
The protomatrix of the SC-PLDPCH-TDC ${\bm{B}_{\rm SC-PLDPCH-TDC}} = \{b_{\rm TDC}(i, j)\}$ therefore has a size of $m(L+W) \times nL$.
In \eqref{eq:example_td}, the protomatrix $\bm{B}$ of size $m \times n = 3 \times 4$ is split into $\bm{B}_0$ and $\bm{B}_1$ of the same size assuming $W=1$.
Based on $\bm{B}_0$ and $\bm{B}_1$, {\color{black} an} SC-PLDPCH-TDC with $W=1$ and $L=2$ is constructed in \eqref{eq:example_B_TDC} and has a size of $m(L+W) \times nL = 9 \times 8$.
Note that in computing the threshold of the SC-PLDPCH-TDC, a larger $L$, e.g., $L=10$, will be used for threshold evaluation such that $nL > m(L+W)$.

Similar to the decoding strategy in the layered decoder \cite{Zhang2021layer}, the layered PEXIT method proceeds as follows.
\begin{enumerate}
\item Set the initial  $E_b/N_0$ in dB (i.e., $E_b/N_0({\rm dB})$) \footnote{The initial values will be different according to different codes (corresponding to different Hadamard order $r$). In our optimization design, given $W=1$ and $L = 10$, initial $E_b/N_0$ is $-0.30$ dB for $r = 4$,  $-0.40$ dB for $r = 5$,  $-0.80$ dB for $r = 8$ and  $-0.85$ dB for $r = 10$, respectively. }.
\item Set the maximum number of iterations $N_{\rm max}=150$.
\item \label{step:cpt_ch} Compute $\sigma_{L_{ch}} = {(8R \cdot 10^{(E_b/N_0({\rm dB})) / 10})}^{1/2}$ {\color{black} for ${L_{ch}}$}, and set $\sigma_{app}(j) = \sigma_{L_{ch}}$ for $j=1,2,\ldots,nL$.
\item For $ i=1,2,\ldots,m(L+W) $ and $ j=1,2,\ldots,nL$, set
${I_{av}}\left( {i,j} \right)=0$.
\item Set the iteration number $N_{\rm it}=1$.
\item \label{step:start} Set $i = 1$.
\item \label{step:sigma_app} For $ j=1,2,\ldots,nL$, subtract $b_{\rm TDC}(i,j) \times {{\left( {{J^{ - 1}}\left( {{I_{av}}\left( {i,j} \right)} \right)} \right)}^2}$ from $({\sigma _{app}}\left( j \right))^2$, and then compute ${\sigma _{temp}}\left( j \right)$ using \eqref{eq:sigma_app_1}.
\item \label{step:Iev}  For $ j=1,2,\ldots,nL$, compute \eqref{eq:I_ev}
 if  $b_{\rm TDC}{(i,j)} > 0$;
 otherwise set  $I_{ev}({i,j}) = 0$.

Taking the first row of the $9 \times 8$ protomatrix $\bm{B}_{\rm SC-PLDPCH-TDC}$ in \eqref{eq:example_B_TDC} as an example, we obtain the $1 \times 8$ vector
 ${\bm{b}_{\rm TDC}(1,:)}$ shown in \eqref{eq:example_1}.
After analyzing the MI of the P-VNs, the corresponding $1 \times 8$ MI vector $\bm{I}_{ev}(1, :)$ is shown in \eqref{eq:I_ev_ex_1}.
\item Convert the $1 \times nL$ $\bm{I}_{ev}(i, :)$ MI vector into a $1 \times d$ $\bm{I}_{ah}$ MI vector by eliminating the $0$ entries and repeating $b_{\rm TDC}(i,j) (\ge 1)$ times
the entry $I_{ev}(i, j)$ .

\textit{Remark:} Each row weight of a protomatrix $\bm{B}$ for the underlying PLDPCH-BC equals $d = r + 2$ \cite{zhang2021} and hence each row corresponds to a $r = d - 2$ Hadamard code.
However, due to the structure of SC-PLDPCH-TDCs, the first and last $Wm$ rows in $\bm{B}_{\rm SC-PLDPCH-TDC}$ contain only part of the structure $[\bm{B}_W\ \ldots\ \bm{B}_1\ \bm{B}_0]$ ($\bm{B} = \sum_{i = 0}^W {{\bm{B}_i}}$) and thus the row weight could be less than $d$.
For simplicity and uniformity, we compute the MI values of $r = d - 2$ Hadamard code for each row (i.e., each H-CN).
For the first/last $Wm$ rows in $\bm{B}_{\rm SC-PLDPCH-TDC}$, when their row weight $d_1$ is less than $d$, the first/last $d-d_1$ MI values in $\bm{I}_{ah}$ are set to $1$ \footnote{MI value equal to 1 means that the corresponding MI is known and does not provide any new information in the analysis. }.
For example,
the row weight of the  underlying protomatrix $\bm{B}$ \eqref{eq:example_td} equals $d=6$ while the row weight of the $1 \times 8$ $\bm{b}_{TDC}(1,:)$ equals $d_1 = 3 < d$.
Hence, when converting $1 \times 8$ $\bm{I}_{ev}(1,:)$ into the $1 \times 6$ $\bm{I}_{ah}$ MI vector, the first $d-d_1 = 3$ values, i.e., $[I_{ah}(1)\ I_{ah}(2)\ I_{ah}(3)]$ are set to $1$, as shown in \eqref{eq:I_ah_1}.

\item Based on $\sigma_{L_{ch}}$ and the $d$ entries in $\bm{I}_{ah}$,
we  use the Monte Carlo method in \cite{zhang2021} to generate $d$ extrinsic MI values, i.e., a $1 \times d$ MI vector $\bm{I}_{eh}$ \footnote{When MI value equals $1$, Monte Carlo method will generate the corresponding LLR values with large absolute values. In our analysis, we set them as $\pm 100$. }.
(For more details of the method, please refer to the Appendices in \cite{zhang2021}.)
We use the same symbol definitions as in \cite{zhang2021}.
Hence $p_e( {\xi |X = x} )$  in \eqref{I_eh_k_old} denotes the PDF of the LLR values given
the bit $x$ being ``$0$'' or ``$1$'';   in \eqref{I_eh_k},
$\bm{U} = \{U(\alpha, k)\}$ denotes a $w \times d$ matrix in which each row represents a length-$d$ SPC codeword and
$\bm{V} = \{V(\alpha, k)\}$ denotes a $w \times d$ matrix in which each row represents a set of ($d$) extrinsic LLR values generated by the Hadamard decoder.

Using the previous example, the MI vector $\bm{I}_{eh}$ is shown in \eqref{I_eh_1}.
\item   Convert the $1 \times d$ $\bm{I}_{eh}$ MI vector into a $1 \times nL$ $\bm{I}_{av}(i,:)$ MI vector.
For $ j=1,2,\ldots,nL$,
if $b_{\rm TDC}(i,j) > 0$, set the value of $I_{av}(i, j)$ as the average of the corresponding $b_{\rm TDC}(i,j)$ MI values in $\bm{I}_{eh}$;
otherwise set $I_{av}(i, j)=0$.
For the first/last $Wm$ rows in $\bm{B}_{\rm SC-PLDPCH-TDC}$, when their row weight $d_1$ is less than $d$, the first/last $d-d_1$ MI values in $\bm{I}_{eh}$ will be omitted, and the remaining $d_1$ values will be used to compute $\bm{I}_{av}(i,:)$.

In the above example, we omit the first $d - d_1 = 3$ MI values in $\bm{I}_{eh}$, i.e, $[I_{eh}(1)\ I_{eh}(2)\ I_{eh}(3)]$, and use the remaining $d_1 = 3$ MI values, i.e., $[I_{eh}(4)\ I_{eh}(5)\ I_{eh}(6)]$ to compute the $\bm{I}_{av}(1,:)$ MI vector given in \eqref{eq:I_av_1}.
\item \label{step:update_APP}  For $ j=1,2,\ldots,nL$, use the ``new'' extrinsic information $\bm{I}_{av}(i,:)$ to update the \emph{a posteriori} information ${\sigma _{app}}(j)$ by \eqref{eq:sigma_app_2}.

\item \label{step:end}  If $i$ is smaller than the number of rows ($m(L+W)$), 
increment $i$ by 1 and go to Step \ref{step:sigma_app}).
\item For $ j=1,2,\ldots,nL$, compute $I_{app}(j) = J(\sigma_{app}(j))$.
If $I_{app}(j)=1 \ \forall \ j$, decrement $E_b/N_0$ by $0.05$ dB and go to Step \ref{step:cpt_ch});
\item If the maximum number of iterations is not reached, increment 
$N_{\rm it}$ by $1$ and go to Step \ref{step:start}); otherwise, stop and the threshold is given by $(E_b/N_0)^* = E_b/N_0 + 0.05$ dB.
\end{enumerate}

As can be observed,
the proposed layered PEXIT chart algorithm updates the corresponding \emph{a posteriori} information (using Step \ref{step:update_APP})) every time the extrinsic MI of one row (layer) is obtained.
Further, we set $L = 10$ and construct {\color{black} an} SC-PLDPCH-TDC with a protomatrix $\bm{B}_{\rm SC-PLDPCH-TDC}$ of size $77\times 110$ (where $\bm{B}_0$  and $\bm{B}_1$ are given by \eqref{mat:th_r4_b0}).
Table \ref{tb:nit} lists the number of iterations required for the
original PEXIT algorithm \cite{zhang2021} and our proposed layered PEXIT algorithm to converge as $E_b/N_0$ changes from $-0.30$ dB to $-0.40$ dB.
We observe that our layered PEXIT algorithm reduces the number of iterations by at least $30$\% compared with the original PEXIT algorithm.
The number of iterations is reduced by $62.8$\% at $E_b/N_0=-0.40$ dB.
We also find that both algorithms do not converge at $E_b/N_0=-0.45$ dB
when the maximum number of iterations  is set to $N_{\rm max} = 150$ for the layered PEXIT chart method, and is set to $2N_{\rm max} = 300$ for original PEXIT chart method.
Thus both algorithms arrive at the same decoding threshold, i.e., $(E_b/N_0)^*=-0.40$ dB.
The proposed layered PEXIT algorithm can therefore speed up the convergence rate when calculating the threshold of SC-PLDPCH-TDCs. 

\begin{table}[t]
\newcommand{\tabincell}[2]{\begin{tabular}{@{}#1@{}}#2\end{tabular}}
\centering\caption{Number of iterations $N_{\rm it}$ required for the PEXIT algorithms to converge at different $E_b/N_0$. The maximum number of iterations is $N_{\rm max}=150$ for the layered PEXIT chart algorithm and $2N_{\rm max}=300$ for the original PEXIT chart algorithm.}\label{tb:nit} \small
\begin{center} 
 \begin{tabular}{|c|c|c|c|}
\hline
$E_b/N_0$ in dB & $-0.30$  & $-0.35$  & $-0.40$   \\
\hline
Original PEXIT \cite{zhang2021} & $107$  & $190$  & $197 $  \\
\hline
Proposed layered PEXIT & $80 $  & $104 $  & $121 $  \\
\hline
\end{tabular}
\end{center}
\end{table}
\textit{Remark}:
The difference between our proposed layered PEXIT algorithm and the shuffled EXIT algorithm \cite{Yang2020} are as follows.
\begin{itemize}
\item Given a protomatrix, our algorithm performs the analysis row by row, while the shuffled EXIT algorithm \cite{Yang2020} performs the analysis column by column.
\item When analyzing SC-PLDPCH-TDCs, our algorithm computes MI values for Hadamard check nodes, while \cite{Yang2020} computes MI values for SPC-CNs. 
\end{itemize}

\subsection{Optimizing Protomatrices using Genetic Algorithm}\label{sect:GA}
To solve a problem based on GA  \cite{Holland1975}, a generation group is first created or initialized
and a fitness function is developed to calculate the fitness value of each individual in the group.
Based on the fitness values, some individuals are selected from the ``parent'' generation group to form an ``offspring'' generation group.
To facilitate obtaining good solutions, the individuals having the best fitness values in the parent generation group will be kept in the offspring generation group.
Crossover and mutation operations are further performed in the offspring generation group.
Subsequently, the ``offspring'' generation group becomes the ``parent'' generation group. By
repeating the generation cycles, GA has a high probability of arriving at the global optimal solution to the problem \cite{Srinivas1994}. 
Given a protomatrix $\bm{B}$ corresponding to a PLDPCH-BC,
we propose a GA to systematically search for optimized sets of
protomatrices $\{ \bm{B}_0, \bm{B}_1, \ldots, \bm{B}_W \}$ (where
$\bm{B} = \sum_{i = 0}^W {{\bm{B}_i}}$) for the corresponding SC-PLDPCH-TDC.
By selecting SC-PLDPCH-TDCs with good thresholds, our final objective is to design optimal SC-PLDPCH-CCs based on the same set of protomatrices.
We denote
\begin{itemize}
\item $K$ parent individuals $\Phi^k$ ($k = 1, 2, \ldots, K$)
as $K$ sets of $W+1$ protomatrices, i.e.,
 $\Phi^k = \{\bm{B}_0^k, \bm{B}_1^k, \ldots, \bm{B}_W^k\}$, each of which satisfies $\bm{B} = \sum_{i = 0}^W {{\bm{B}_i^k}}$;
\item the fitness function as $ \psi(\cdot)$
and the fitness value of the $k$-th parent individual as $f_k = \psi(\Phi^k)$;
\item the probability of selecting the $k$-th parent individual as ${p_{s_k}}$;
\item the probability of crossover as $p_c$ and
the probability of mutation as $p_m$;
\item $K$ offspring individuals $\Upsilon_k$ $(k = 1, 2, \ldots, K)$
as $K$ sets of  $W+1$ protomatrices, i.e.,
$\Upsilon_k= \{ \bm{S}_0^k, \bm{S}_1^k, \cdots, \bm{S}_W^k\}$, each of which satisfies $\bm{B} = \sum_{i = 0}^W {{\bm{S}_i^k}}$;
\item $m \times n$ as the size of $\bm{B}$, and hence also the size of $\bm{B}_i^k$
and $\bm{S}_i^k$ for all $i$ and $k$;
\item $N_{g}$ as the number of individuals having good fitness values.
\end{itemize}
Our proposed GA algorithm is described as follows.
\begin{enumerate}
\item Initialization: Set coupling width $W$, coupling length $L$, and the number of individuals $K$ in each generation.
Randomly generate
$K$ sets of  $\Phi^k = \{\bm{B}_0^k, \bm{B}_1^k, \ldots, \bm{B}_W^k\}$, $k = 1, 2, \ldots, K$.
\item \label{step:start_b} Computation of fitness values: For each $\Phi^k$, we
construct the corresponding SC-PLDPCH-TDC by coupling $L$ sets of $\Phi^k$.
Then we apply our proposed layered PEXIT chart method in Section~\ref{sect:layered_PEXIT}  to analyze the threshold
and convergence behavior of the SC-PLDPCH-TDC.
Recall that for a specific $E_b/N_0$ value (in dB),
 the iteration number required to converge is given by $N_{\rm it}$
and the maximum number of iterations is given by $N_{\rm max}$.
Our proposed fitness function $\psi(\cdot)$ takes both
the number of successful convergences, denoted by $N_c$, and
the convergence rate  $N_{\rm it}$ into consideration.
We define $\psi(\cdot)$ as
\small
 \begin{equation}\label{eq:fit_func}
 \psi(\Phi^k) =f_k= \sum_{i = 1}^{{N_c}} \left[ N_{\rm max} - N^k_{{\rm it},i} \right]
= {N_c} {N_{\rm max}} - \sum_{i = 1}^{{N_c}} {{N^k_{{\rm it},i}}}
\end{equation}
\normalsize
where  ${N^k_{{\rm it},i}}$ represents the number of iterations for the
SC-PLDPCH-TDC corresponding to $\Phi^k$ to converge in the
$i$-th successful convergence ($i=1,2,\ldots,N_c$).
 Note that $\psi(\Phi^k)$ should give a larger value if the
 corresponding SC-PLDPCH-TDC converges faster and more times
 in the layered PEXIT algorithm.

In the example given in Table \ref{tb:nit},
the layered PEXIT algorithm has converged at
 $-0.30$ dB, $-0.35$ dB and $-0.40$ dB, and hence
 $N_c = 3$.
 Using \eqref{eq:fit_func}, the fitness value of $\{ {\bm{B}_0}, {\bm{B}_1} \}$
 defined in \eqref{mat:th_r4_b0} 
 is therefore given by
$\psi(\{ {\bm{B}_0}, {\bm{B}_1} \}) = {3} \times {150} - (80+104+121) = 145.$

\item Selection:
Compute the fitness values $f_k= \psi(\Phi^k) $ for the $K$ parent individuals $\Phi^k$ ($k = 1, 2,\ldots,K$).
Subsequently, we normalize $f_{k}$ ($k = 1, 2,\ldots\ K$) to form the probabilities of selection, i.e.,
${p_{s_k}} = {f_{k}}/\sum\limits_{j = 1}^K {{f_j}};\ k = 1, 2,\ldots\ K$.
Then the $K$ offspring individuals $\Upsilon_{k'}$ $(k' = 1, 2, \ldots, K)$
are chosen from the parent generation group
$\Phi^k$ ($k = 1, 2, \ldots, K$) as follows.
First, the $N_g$ individuals with the highest fitness values in the parent group are passed to the offspring group directly to fill $\Upsilon_{k'}$ $(k' = 1, 2, \ldots, N_g)$.
Then to fill each of the remaining $K- N_g$ offspring slots, i.e., $\Upsilon_{k'}$ $(k' = N_g+1, N_g+2, \ldots, K)$, a random parent individual $\Phi^k$ is selected accordingly to the
probability ${p_{s_k}}$. These $K- N_g$ offsprings will further go through the crossover and mutation processes below.

\item Crossover:
Assuming $K-N_g$ is even,
the $K-N_g$ selected offspring individuals, i.e.,  $\Upsilon_{k'}$ $(k' = N_g+1, N_g+2, \ldots, K)$
are randomly divided into $(K-N_g)/2$ pairs.
For each pair
$\Upsilon_{k_1}= \{ \bm{S}_0^{k_1}, \bm{S}_1^{k_1}, \cdots, \bm{S}_W^{k_1}\}$
and $\Upsilon_{k_2}= \{ \bm{S}_0^{k_2}, \bm{S}_1^{k_2}, \cdots, \bm{S}_W^{k_2}\}$,
we select two random positions $(u_1,v_1)$ and $(u_2,v_2)$
where $1 \le u_1 \le u_2 \le m$ and $1 \le v_1 \le v_2 \le n$.
With a probability of $p_c$, all entries between
$(u_1,v_1)$ and $(u_2,v_2)$ (in a column-wise manner) in $\bm{S}_i^{k_1}$
are exchanged with the corresponding entries in $\bm{S}_i^{k_2}$
 $(i = 0,1, \ldots, W)$.

Fig. \ref{fig:crossover_mutation}(a) shows a crossover example
where $m \times n = 3 \times 4$.
The pair of offspring individuals selected are given by
$\Upsilon_{k_1}= \{ \bm{S}_0^{k_1}, \bm{S}_1^{k_1}\}$
and $\Upsilon_{k_2}=\{ \bm{S}_0^{k_2}, \bm{S}_1^{k_2}\}$ ($N_g < k_1 \ne k_2 \le K$).
The two selected positions are $(u_1,v_1)=(2,2)$ and $(u_2,v_2)=(2,3)$.
Hence the entries in positions $(2,2), (3,2),(1,3),(2,3)$
are exchanged between $\bm{S}_0^{k_1}$ and $\bm{S}_0^{k_2}$;
and are exchanged between $\bm{S}_1^{k_1}$ and $\bm{S}_1^{k_2}$.

\item Mutation: The offspring individuals
$\Upsilon_{k'}= \{ \bm{S}_0^{k'}, \bm{S}_1^{k'}, \cdots, \bm{S}_W^{k'}\}$
that have gone through the crossover process
will then be considered for mutation separately.
Within each offspring individual, one random position $(u,v)$ $(1 \le u \le m$ and $1 \le v \le n)$ corresponding to a non-zero entry in $\bm{B}$ are selected.
For all the $(u,v)$-th entries in $\bm{S}_0^{k'}, \bm{S}_1^{k'}, \cdots, \bm{S}_W^{k'}$, they are mutated together with a probability of $p_m$. The entries should have different values after
the mutation while keeping  $ \sum_{i = 0}^W {{\bm{S}_i^{k'}}}=\bm{B}$ satisfied.

Fig. \ref{fig:crossover_mutation}(b) illustrates a mutation example.
The offspring individual being considered is
$\Upsilon_{k_1}= \{ \bm{S}_0^{k_1}, \bm{S}_1^{k_1}\}$ ($N_g < k_1 \le K$)
and mutation occurs in all the $(3,1)$-th entries, where the $(3,1)$-th entry in $\bm{B} = \bm{S}_0^{k_1} + \bm{S}_1^{k_1}$ \eqref{eq:example_td} equals $3$ and hence is non-zero.

\item \label{step:end_b} Set $\Phi_k = \Upsilon_k$, $k = 1, 2, \dots, K$.
\item Repeat Steps \ref{step:start_b}) to \ref{step:end_b}) until
$\exists \Phi_k$ such that
 the corresponding SC-PLDPCH-TDC has a low threshold and
the derived 
SC-PLDPCH-CC has a good error performance at low $E_b/N_0$.
\end{enumerate}

\begin{figure}[t]
\centerline{\includegraphics[width=0.5\columnwidth]{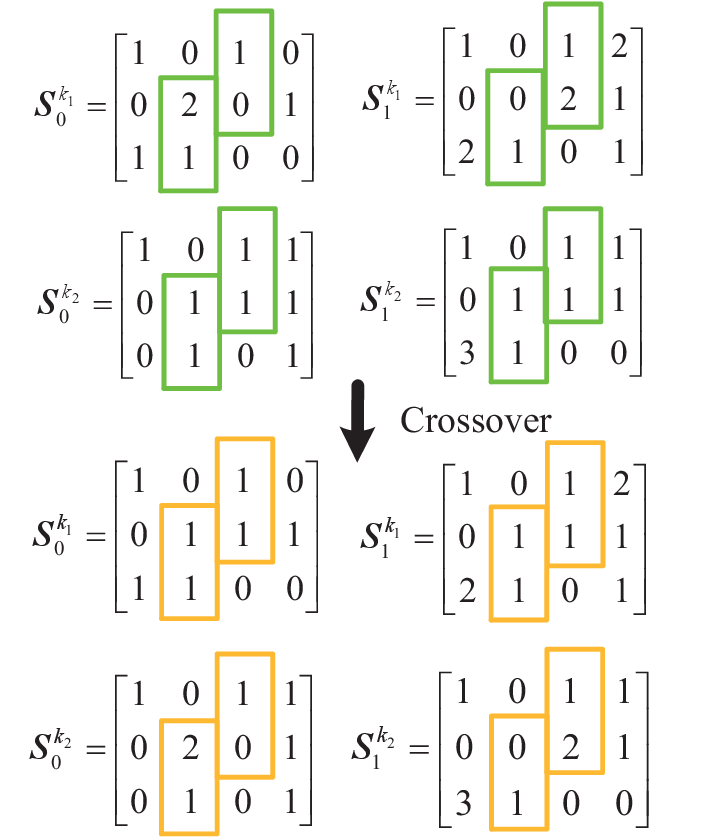} }
\centerline{(a)} 
\vspace{0.2cm}
\centerline{\includegraphics[width=0.5\columnwidth]{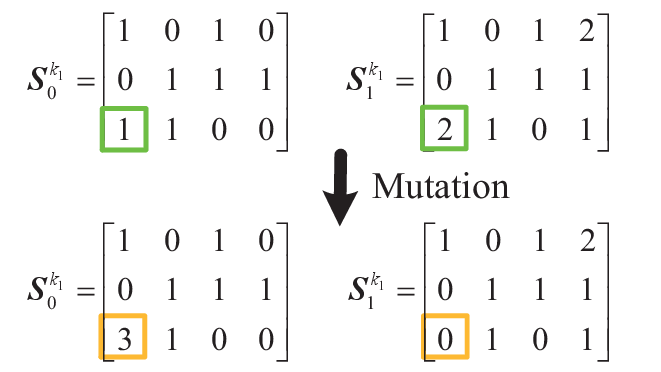} }
\centerline{(b)}
\caption{(a) Crossover example.
The pair of offspring individuals are
$\Upsilon_{k_1}= \{ \bm{S}_0^{k_1}, \bm{S}_1^{k_1}\}$
and $\Upsilon_{k_2}=\{ \bm{S}_0^{k_2}, \bm{S}_1^{k_2}\}$ ($N_g < k_1 \ne k_2 \le K$).
Two selected positions are $(u_1,v_1)=(2,2)$ and $(u_2,v_2)=(2,3)$.
The entries in positions $(2,2), (3,2),(1,3),(2,3)$ of  $\bm{S}_0^{k_1}$, i.e., $[2\ 1\ 1\ 0]$
are exchanged with those in the same positions of $\bm{S}_0^{k_2}$, i.e., $[1\ 1\ 1\ 1]$.
Similarly, $[0\ 1\ 1\ 2]$ in $\bm{S}_1^{k_1}$ are exchanged with $[1\ 1\ 1\ 1]$ in $\bm{S}_1^{k_2}$. 
(b) Mutation example.
$\Upsilon_{1}= \{ \bm{S}_0^{k_1}, \bm{S}_1^{k_1}\}$ ($N_g < k_1 \le K$)
and mutation occurs in all the $(3,1)$-th entries.
The entry $1$ in $\bm{S}_0^{k_1}$ is changed into $3$ and the entry $2$ in $\bm{S}_1^{k_1}$ is changed into $0$ to keep ${\bm{S}_0^{k_1}} + {\bm{S}_1^{k_1}} = \bm{B}$ \eqref{eq:example_td}.
}
  \label{fig:crossover_mutation}
\end{figure}
%


When GA is used to systematically search for   {\color{black}good protomatrices}, setting $L=10$ in the layered PEXIT algorithm can efficiently calculate the threshold of {\color{black} an} SC-PLDPCH-TDC. Once  {\color{black}good protomatrices} are found, we increase $L$ so that the code rate of the SC-PLDPCH-TDC is very close to that of its SC-PLDPCH-CC.
We then use the layered PEXIT method to calculate the threshold of the lengthened SC-PLDPCH-TDC as proxy for the SC-PLDPCH-CC threshold.

\begin{table}[t]
\centering\caption{Thresholds for SC-PLDPCH-TDCs and PLDPCH-BCs \cite{zhang2021} computed by the proposed PEXIT algorithm. $N_{\rm max}^{'}=1000$. For SC-PLDPCH-TDCs, coupling width $W = 1$ and coupling length $L=500$ for $r = 4$;  $L=300$ for $r = 5,8$; and $L=100$ for $r = 10$. }\label{tb:threshold} 
\begin{center}
\begin{tabular}{|c|c|c|c|c|}
\hline
\multicolumn{2}{|c|}{Code} & TDC & BC \cite{zhang2021}  \\ \hline \hline
\multirow{2}{*}{$r=4$}  & Code rate $R$ & $0.0491$  & $0.0494 $  \\ \cline{2-4}
                      &Threshold in dB & $-1.35 $  & $-1.34$ \\ \hline\hline
\multirow{2}{*}{$r=5$}  & Code rate $R$ & $0.021$  & $0.021$  \\ \cline{2-4}
                      &Threshold in dB &  $-1.37$  &  $-1.37$ \\ \hline\hline
\multirow{2}{*}{$r=8$}  & Code rate $R$ &  $0.008$  & $0.008$  \\ \cline{2-4}
                      &Threshold in dB &  $-1.45 $  &  $-1.44$  \\ \hline\hline
\multirow{2}{*}{$r=10$}  & Code rate $R$ & $0.00291$  & $0.00295$  \\ \cline{2-4}
                      &Threshold in dB & $-1.48$  &  $-1.50$  \\ \hline
\end{tabular}
\end{center}
\end{table}

\section{Simulation Results}\label{sect:sim_re}
Based on the optimized PLDPCH-BCs found in \cite{zhang2021},
we search for good SC-PLDPCH-CCs using the layered PEXIT algorithm and the GA
proposed in Section~\ref{sect:layered_PEXIT} and Section~\ref{sect:GA}.  
We assume $W = 1$, and use $L = 10$ and $N_{\rm max} = 150$ in the layered PEXIT method.
We also set $K = 30$, $N_g = 4$, $p_c = 0.8$ and $p_m = 0.6$ in the GA.
We use binary phase-shift-keying (BPSK) modulation over an AWGN channel.
Moreover, we apply the pipeline decoding with different number of processors to evaluate the error performance of the  SC-PLDPCH-CCs found.

For each $E_b/N_0$ value, simulations are performed for at least $1000$ sub-blocks with at least $100$ sub-block errors. Then the corresponding BER is evaluated.
Under comparable information lengths and code rates, we compare the BER performance
of our proposed codes not only with that of the underlying PLDPCH-BC,
but also with those of state-of-the-art codes, namely
LDPC-Hadamard block code (LDPCH-BC)  \cite{Yue2007}, turbo-Hadamard code (THC) \cite{Li2003},
irregular zigzag-Hadamard code (IRZHC) \cite{Li2008,Li2008b} (whenever BER results are available).
These long codes with good error performance under very low $E_b/N_0$ values
can potentially be applied to deep space communications where the signal from the space probe
to Earth is extremely weak; and to embed/transmit hidden information
over an ordinary wireless communication link.
{\color{black} 
For $r=4$ and $r=5$, we also compare the BER performance of PLDPCH-BCs and SC-PLDPCH-CCs under the same blocklength/constraint length.}

\begin{figure*}[!t]
\begin{equation}\label{mat:r4_b0}{
{\bm{B}_0} = \left[ {\begin{array}{*{20}{c}}
1&    0&    0&    0&    0&    0&    1&    0&    1&    0&    0\\
0&    0&    1&    0&    0&    0&    0&    0&    0&    2&    1\\
2&    1&    0&    0&    0&    0&    0&    0&    0&    0&    1\\
0&    1&    0&    1&    0&    0&    0&    0&    0&    2&    0\\
1&    0&    0&    0&    0&    0&    0&    1&    0&    1&    0\\
1&    0&    0&    2&    0&    0&    0&    0&    0&    0&    0\\
1&    0&    0&    0&    1&    0&    0&    0&    0&    2&    0\\
\end{array}} \right] \;\;\;
{\bm{B}_1} = \left[ {\begin{array}{*{20}{c}}
0&    0&    0&    0&    0&    0&    0&    0&    2&    0&    1\\
0&    1&    1&    0&    0&    0&    0&    0&    0&    0&    0\\
0&    0&    0&    0&    1&    1&    0&    0&    0&    0&    0\\
0&    0&    0&    2&    0&    0&    0&    0&    0&    0&    0\\
1&    0&    0&    0&    0&    0&    0&    0&    0&    2&    0\\
2&    0&    0&    0&    0&    0&    1&    0&    0&    0&    0\\
0&    0&    0&    1&    0&    0&    0&    0&    1&    0&    0
\end{array}} \right]}
\end{equation}
\begin{equation}\label{mat:r5_b0}{
{\bm{B}_0} = \left[ {\begin{array}{*{20}{c}}
1&    2&    0&    0&    0&    0&    0&    0&    1&    0\\
0&    0&    1&    0&    0&    1&    0&    2&    0&    0\\
0&    0&    0&    1&    1&    0&    0&    0&    0&    1\\
0&    0&    0&    1&    0&    0&    0&    2&    0&    1\\
0&    0&    0&    1&    0&    0&    1&    2&    0&    0\\
1&    0&    0&    0&    0&    0&    0&    1&    0&    1\\
\end{array}} \right] \;\;\;
{\bm{B}_1} = \left[ {\begin{array}{*{20}{c}}
2&    0&    0&    0&    1&    0&    0&    0&    0&    0\\
0&    0&    1&    0&    0&    1&    1&    0&    0&    0\\
0&    0&    0&    2&    0&    0&    0&    1&    0&    1\\
0&    1&    0&    0&    0&    0&    0&    0&    0&    2\\
0&    0&    0&    1&    0&    0&    0&    0&    0&    2\\
1&    0&    1&    1&    0&    0&    0&    1&    0&    0
\end{array}} \right]}
\end{equation}
\begin{equation}\label{mat:r8_b0}
{
{\bm{B}_0} = \left[ {\begin{array}{*{20}{c}}
1&    0&    1&    0&    0&    0&    0&    1&    0&    0&    0&    1&    0&    0&    0\\
0&    1&    0&    0&    1&    0&    0&    0&    0&    0&    0&    0&    0&    2&    0\\
0&    0&    0&    0&    0&    0&    1&    0&    0&    1&    1&    2&    1&    0&    0\\
0&    0&    0&    1&    2&    0&    0&    0&    0&    0&    0&    3&    0&    0&    1\\
0&    0&    0&    0&    0&    1&    0&    0&    1&    1&    0&    0&    1&    0&    0\\
\end{array}} \right]
}
\end{equation}
\begin{equation}\label{mat:r8_b1}
{
{\bm{B}_1} = \left[ {\begin{array}{*{20}{c}}
1&    0&    0&    0&    0&    0&    0&    2&    2&    0&    0&    0&    0&    0&    1\\
0&    1&    0&    1&    0&    0&    0&    0&    0&    0&    0&    3&    0&    1&    0\\
0&    0&    1&    0&    0&    2&    1&    0&    0&    0&    0&    0&    0&    0&    0\\
0&    0&    0&    1&    0&    0&    0&    0&    0&    1&    0&    0&    0&    0&    1\\
0&    0&    0&    0&    0&    0&    1&    0&    0&    0&    1&    2&    2&    0&    0
\end{array}} \right]}.
\end{equation}
\begin{equation}\label{mat:r10_b0}{
{\bm{B}_0} = \left[ {\begin{array}{*{24}{c}}
1&    0&    0&    0&    0&    0&    2&    0&    0&    0&    0&    0&    0&    0&    1&    0&    0&    0&    0&    1&    3&    0&    1&    0\\
0&    0&    0&    1&    1&    0&    0&    0&    1&    1&    0&    0&    1&    0&    0&    0&    2&    0&    0&    0&    0&    0&    0&    0\\
0&    1&    1&    0&    0&    1&    0&    0&    0&    0&    0&    0&    0&    0&    0&    0&    0&    0&    2&    0&    0&    0&    0&    0\\
0&    0&    0&    0&    0&    0&    0&    1&    0&    0&    1&    1&    0&    0&    0&    0&    0&    0&    0&    0&    1&    1&    0&    0\\
0&    1&    0&    0&    0&    0&    0&    0&    0&    0&    0&    0&    0&    0&    0&    1&    0&    1&    0&    2&    0&    0&    1&    0\\
0&    0&    0&    0&    0&    0&    0&    2&    2&    2&    0&    0&    0&    1&    0&    0&    0&    0&    0&    0&    0&    0&    0&    0\\
\end{array}} \right]}
\end{equation}
\begin{equation}\label{mat:r10_b1}{
{\bm{B}_1} = \left[ {\begin{array}{*{24}{c}}
0&    0&    0&    0&    0&    0&    1&    0&    0&    0&    0&    0&    0&    0&    1&    0&    0&    0&    0&    0&    1&    0&    0&    0\\
0&    0&    0&    2&    1&    0&    0&    0&    0&    0&    0&    0&    0&    0&    0&    0&    1&    1&    0&    0&    0&    0&    0&    0\\
0&    0&    1&    0&    0&    0&    0&    0&    0&    0&    0&    0&    0&    0&    0&    1&    0&    0&    1&    0&    4&    0&    0&    0\\
0&    0&    0&    0&    0&    0&    0&    0&    0&    0&    2&    2&    0&    0&    0&    0&    0&    0&    0&    0&    1&    1&    0&    1\\
2&    1&    0&    0&    0&    0&    0&    0&    0&    0&    0&    0&    0&    1&    0&    0&    0&    0&    0&    1&    0&    0&    0&    1\\
0&    0&    0&    0&    0&    1&    0&    1&    1&    0&    0&    0&    1&    0&    0&    0&    0&    0&    0&    0&    1&    0&    0&    0
\end{array}} \right]}
\end{equation}
 \hrulefill
\end{figure*}

\subsection{Rate-$0.0494$ and $r = 4$}
Based on the $7\times 11$ protomatrix $\bm{B}$ 
of the optimized rate-$0.0494$ PLDPCH-BC \cite{zhang2021}, we apply GA and find two $7\times 11$ protomatrices shown as
\eqref{mat:r4_b0}
after $13$ generations.
The corresponding fitness value equals $554$. Then we increase $L$ to $500$ and the code rate of the SC-PLDPCH-TDC constructed by \eqref{mat:r4_b0}
 is increased to about $0.0491$, approaching that of underlying PLDPCH-BC \cite{zhang2021}.
Using the proposed layered PEXIT method with $N_{max}^{'} = 1000$ iterations, Table \ref{tb:threshold} lists the thresholds for the $r=4$ SC-PLDPCH-TDC with $L=500$, and the $r=4$ PLDPCH-BC  \footnote{ In \cite{zhang2021}, \eqref{I_eh_k_old} has been used to compute
$\bm{I}_{eh}$. It is not very efficient because of the need to evaluate the PDF of the LLR values.
In this paper,  \eqref{I_eh_k} is used instead  to compute
$\bm{I}_{eh}$ because it can be evaluated much more efficiently with graphics processing units.
The computed thresholds are found to be slightly different and larger than those reported in  \cite{zhang2021}.} \cite{zhang2021}, respectively.
The theoretical threshold of the terminated code with $L=500$ is found to be $-1.35$ dB,
which is approximated as the threshold of the SC-PLDPCH-CC constructed by \eqref{mat:r4_b0}. 
Note that the threshold
is slightly lower
{\color{black}(i.e., slightly better)} than that of the PLDPCH-BC.
We use the lifting factors $z_1 = 32$ and $z_2 = 512$ to expand
the protomatrix such that the sub-block length of the SC-PLDPCH-CC equals
$1,327,104$, which is identical to the
 code length of the PLDPCH-BC with $z_1 = 32$ and $z_2 = 512$, i.e., $N+M(2^r -r-2) = 1,327,104$.
 {\color{black} Moreover, the information length in each sub-block/block is $65536$ for both codes.}
The BER performance of the SC-PLDPCH-CC with different number of processors $I$  used in pipeline decoding  is shown
 in Fig. \ref{fig:r4}(a).
We observe that the decoder with $I=80$ processors in pipeline decoding achieves a BER of $10^{-5}$ at about $E_b/N_0 = -1.24$ dB,
 which outperforms that with (i) {\color{black} $I = 75$ by about $0.015$ dB}, (ii) $I = 70$ by about $0.03$ dB, and (iii) $I = 60$ by about $0.06$ dB.
{\color{black} The gaps of the SC-PLDPCH-CC (with $I=80$ and BER of $10^{-5}$) to the PEXIT threshold ($-1.35$ dB)  and to the ultimate Shannon limit ($-1.59$ dB) are about $0.11$ dB and $0.35$ dB, respectively.}

{\color{black}
In the same figure, we plot the BER results of the underlying PLDPCH-BC using {\color{black}$I_{max}=300$} standard decoding iterations, 
which should achieve almost the same BER performance
as  {\color{black}$I_{max}/2 = 150 = I_{BC}$} layered decoding iterations \cite{Zhang2021layer}.
We recall in Sect.~\ref{sect:CA} that when $(W+1) I = I_{BC}$,
the latencies of both SC-PLDPCH-CC pipeline decoder and PLDPCH-BC decoder
become identical.
Thus when $I_{BC}=150$ layered decoding iterations are used in
the PLDPCH-BC decoder, the number of processors used in
the SC-PLDPCH-CC pipeline decoder equals $ I = 150/(W+1) = 75$.
Comparing the corresponding BER curves in the figure shows that the SC-PLDPCH-CC
outperforms the PLDPCH-BC
 by about $0.04$ dB at a BER of $10^{-5}$.
   {\color{black}  {\color{black}Next, we increase the second lifting factor of the PLDPCH-BC 
   by a factor of  $\delta_{z_2}  = 1+ \frac{W}{1 + (2^r-r-2)\frac{m}{n}} = 1.136$
   such that the blocklength of the PLDPCH-BC is the same as constraint length
   of the SC-PLDPCH-CC.
 Thus the second lifting factor of the PLDPCH-BC becomes $z_2' = 512 \times 1.136 \approx 582$.
To maintain the same decoding latency, the number of layered iterations is 
reduced from $I_{BC}=150$ to
$I'_{BC}=I_{BC}/\delta_{z_2}=150/1.136 = 132$. In other words, 
$I_{max}=300$ is reduced to $I'_{max}=300/1.136 \approx 264$.}
Comparing the BER curves in the figure shows that 
under the same constraint length/blocklength
and the same decoding latency, the SC-PLDPCH-CC
outperforms the PLDPCH-BC 
 by about $0.03$ dB at a BER of $10^{-5}$.
  }

{\color{black}
We further compare our proposed code with
the rate-$0.0453$ irregular zigzag-Hadamard code (IRZHC) \cite{Li2008}, the rate-$0.05$ LDPC-Hadamard block code (LDPCH-BC) \cite{Yue2007} and the rate-$0.033$ turbo-Hadamard code (THC) \cite{Li2003,Yue2007}.
To have a fair comparison, the information lengths in each block are about $65536$ for all codes.
First,  it is given in
{\color{black}\cite[Fig.~10]{Yue2007}} that
 the rate-$0.033$ THC (with $80$  iterations) achieves
 a BER of $10^{-5}$  at $E_b/N_0=-0.95$ dB.
  In Fig. \ref{fig:r4}, we further redraw the BER curves of  IRZHC
  {\color{black}(\cite[Fig.~12]{Li2008})}
 and LDPCH-BC
 {\color{black}(\cite[Fig.~10]{Yue2007}).}
As can be observed, IRZHC and LDPCH-BC (with  {\color{black}$I_{max}=400$} standard decoding iterations) achieve
a BER of $10^{-5}$  at $E_b/N_0=-1.17$ dB and $-1.18$ dB, respectively.
However,  our SC-PLDPCH-CC (using pipeline decoder with $80$ processors) can attain the same BER
 at  $E_b/N_0=-1.238$ dB.
}
{\color{black}Finally, we use the lifting factors $z_1 = 32$ and $z_2 = 4$ to expand
the protomatrix such that the sub-block length of the SC-PLDPCH-CC equals
$10368$. Fig. \ref{fig:r4}(b) plots the BER results of the code.
At $I=80$ and a BER of $10^{-5}$,
the SC-PLDPCH-CC degrades from $-1.24$~dB to $-0.85$~dB 
when the sub-block length is reduced from $1,327,104$ to $10368$.
}

\begin{figure}[!ht]
\centerline{\includegraphics[width=0.6\columnwidth]{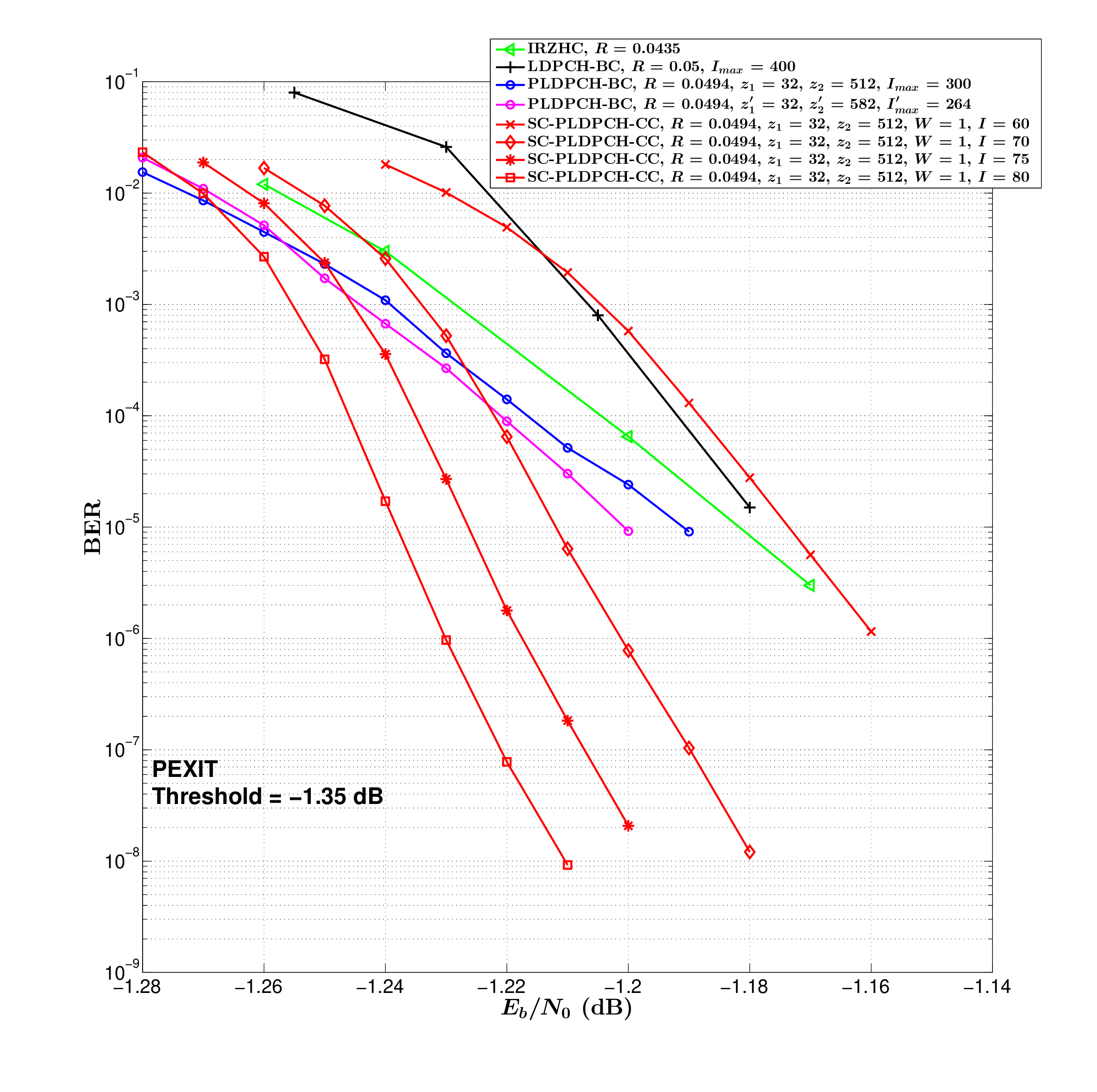}} 
\centerline{
 (a)} 
\centerline{\includegraphics[width=0.6\columnwidth]{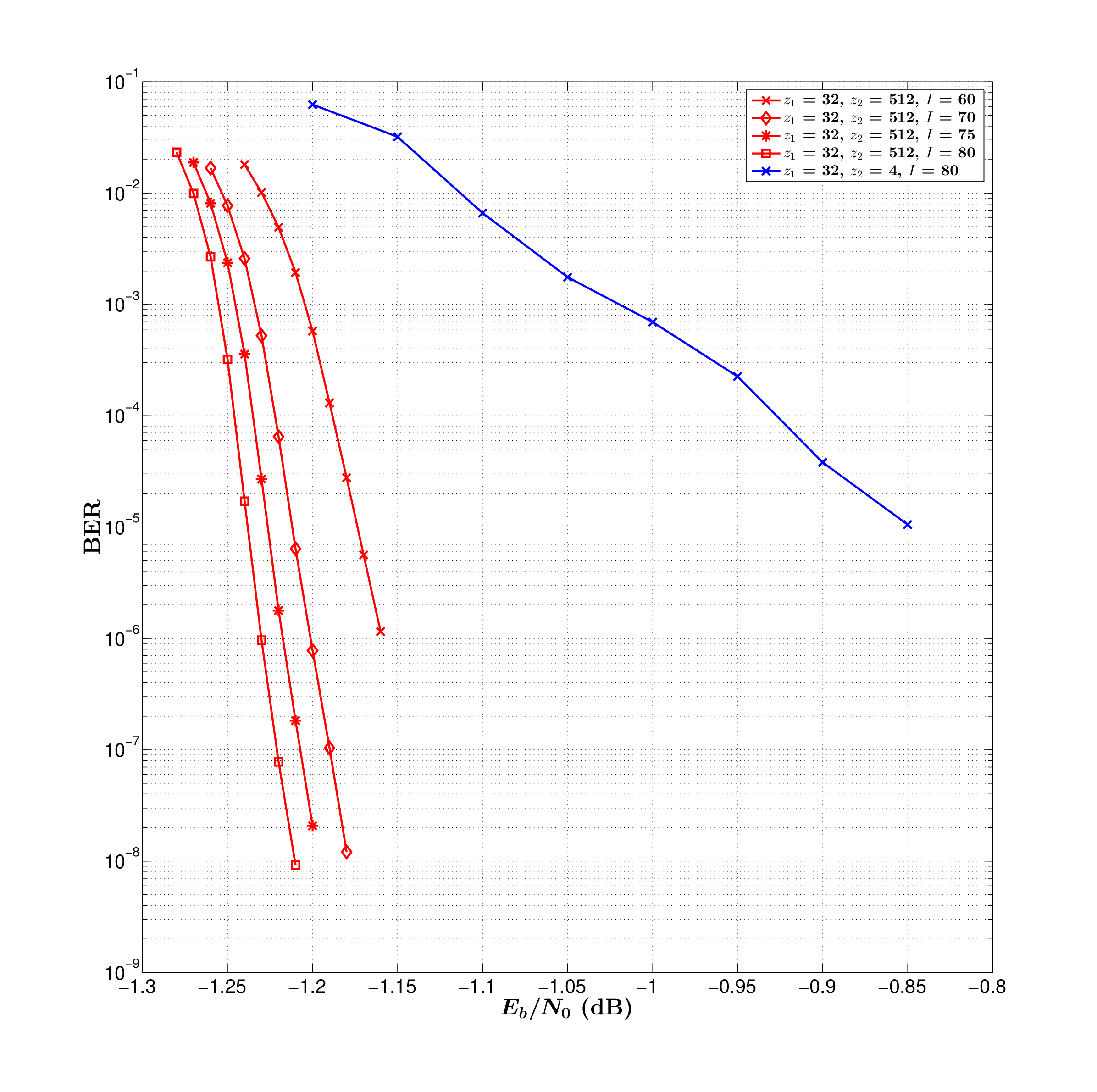}}
\centerline{
 (b)}
\caption{(a) BER performance of rate-$0.0494$ SC-PLDPCH-CC
(PEXIT threshold = $-1.35$ dB), rate-$0.0494$ PLDPCH-BC \cite{zhang2021},
 rate-0.0435 IRZHC \cite{Li2008} and rate-0.05 LDPCH-BC \cite{Yue2007}.
 Information lengths are around $65536$
 for all codes.
 $r = 4$, lifting factors $z_1=32$ and $z_2 = 512$ for SC-PLDPCH-CC.
 $r = 4$, lifting factor sets (i) $z_1=32$ and $z_2 = 512$; 
 and (ii) $z'_1=32$ and $z'_2 = 582$ 
 for PLDPCH-BC. 
(b) BER performance of rate-$0.0494$ SC-PLDPCH-CC with different sub-block lengths. 
}
  \label{fig:r4}
\end{figure}

\subsection{Rate-$0.021$ and $r = 5$}
Based on the $6\times 10$ protomatrix
of the optimized rate-$0.021$ PLDPCH-BC \cite{zhang2021},
 we apply GA and find the $6\times 10$ protomatrices
 shown as \eqref{mat:r5_b0}
after 50 generations.
The corresponding fitness value equals $609$.
Using the proposed layered PEXIT method, Table \ref{tb:threshold} lists the thresholds for the $r = 5$ SC-PLDPCH-TDC constructed by \eqref{mat:r5_b0} and for the $r = 5$ PLDPCH-BC \cite{zhang2021}.
The theoretical threshold of the SC-PLDPCH-TDC with $L = 300$ is estimated
to be $-1.37$ dB, which is the same with that of the PLDPCH-BC \cite{zhang2021}.
Hence, we consider that the $r=5$ SC-PLDPCH-CC constructed by \eqref{mat:r5_b0} has a threshold of about $-1.37$ dB.
We use the same lifting factors, i.e., $z_1 = 32$ and $z_2 = 512$,
as those used in the PLDPCH-BC to expand
the protomatrix such that the sub-block length of the SC-PLDPCH-CC equals
 $3,112,960$.
 {\color{black} The information length in each sub-block/block is $65536$ for both codes.}

The BER performance of the SC-PLDPCH-CC with different $I$ is shown in Fig. \ref{fig:r5}.
The pipeline decoder with $I=80$ processors achieves a BER of $10^{-5}$ at about $E_b/N_0 = -1.30$ dB,
 which outperforms that with (i) {\color{black}$I = 75$ by about $0.01$ dB,}
 (ii) $I = 70$ by about $0.02$ dB,
 and (iii) $I = 60$ by about $0.05$ dB.
{\color{black}  The gaps of the SC-PLDPCH-CC (with $I=80$ and BER of $10^{-5}$) to the PEXIT threshold ($-1.37$ dB) and to the ultimate Shannon limit ($-1.59$ dB) are about $0.07$ dB and $0.29$ dB, respectively.}
In the same figure,
{\color{black}
we  can also observe that under the same latency, SC-PLDPCH-CC using $I = 75$ processors outperforms PLDPCH-BC using $I_{BC}=150$ layered decoding iterations
(equivalent to  {\color{black}$I_{max} = 300$} standard decoding iterations, i.e., the blue curve with symbol ``$\circ$'')
by about  $0.05$ dB at a BER of $10^{-5}$.
}
 {\color{black}  {\color{black} Next, we increase the second lifting factor of the PLDPCH-BC 
   by a factor of  $\delta_{z_2}  = 1+ \frac{W}{1 + (2^r-2)\frac{m}{n}} = 1.053$
   such that the blocklength of the PLDPCH-BC is the same as constraint length
   of the SC-PLDPCH-CC.
 Thus 
 the second lifting factor of the PLDPCH-BC equals $z_2' = 512 \times \delta_{z_2} = 512 \times 1.053 =539$.
To maintain the same decoding latency,
$I'_{BC}=I_{BC}/\delta_{z_2}=150/1.053 \approx 142$ and
$I'_{max}=I_{max}/1.053 \approx 284$.}
Comparing the BER curves in the figure shows that 
under the same constraint length/blocklength
and the same decoding latency, the SC-PLDPCH-CC
outperforms the PLDPCH-BC 
 by more than $0.05$ dB at a BER of $10^{-5}$.
  }

{\color{black}
 In Fig. \ref{fig:r5}, we further redraw the BER curves of
 rate-$0.0189$ THC
 {\color{black}\cite[Fig.~11]{Li2003}, }
 rate-$0.018$ IRZHC
 {\color{black}\cite[Fig.~14]{Li2008b} }
 and rate-$0.022$ LDPCH-BC
 {\color{black}\cite[Fig.~12]{Yue2007}.}
To have a fair comparison, the information lengths in each block are about $65536$ for all codes.
At a BER of $10^{-5}$,
THC, IRZHC (with  {\color{black}$I_{max}=150$} iterations) and LDPCH-BC (with  {\color{black}$I_{max}=400$} iterations)
require $E_b/N_0$ values of $-1.17$ dB, $-1.24$ dB and $-1.26$ dB, respectively; while
our rate-$0.021$ SC-PLDPCH-CC using the pipeline decoding with $I=80$ processors
requires only $E_b/N_0 = -1.30$ dB. 
}

\begin{figure}[!thp]
\centerline{
\includegraphics[width=0.7\columnwidth]{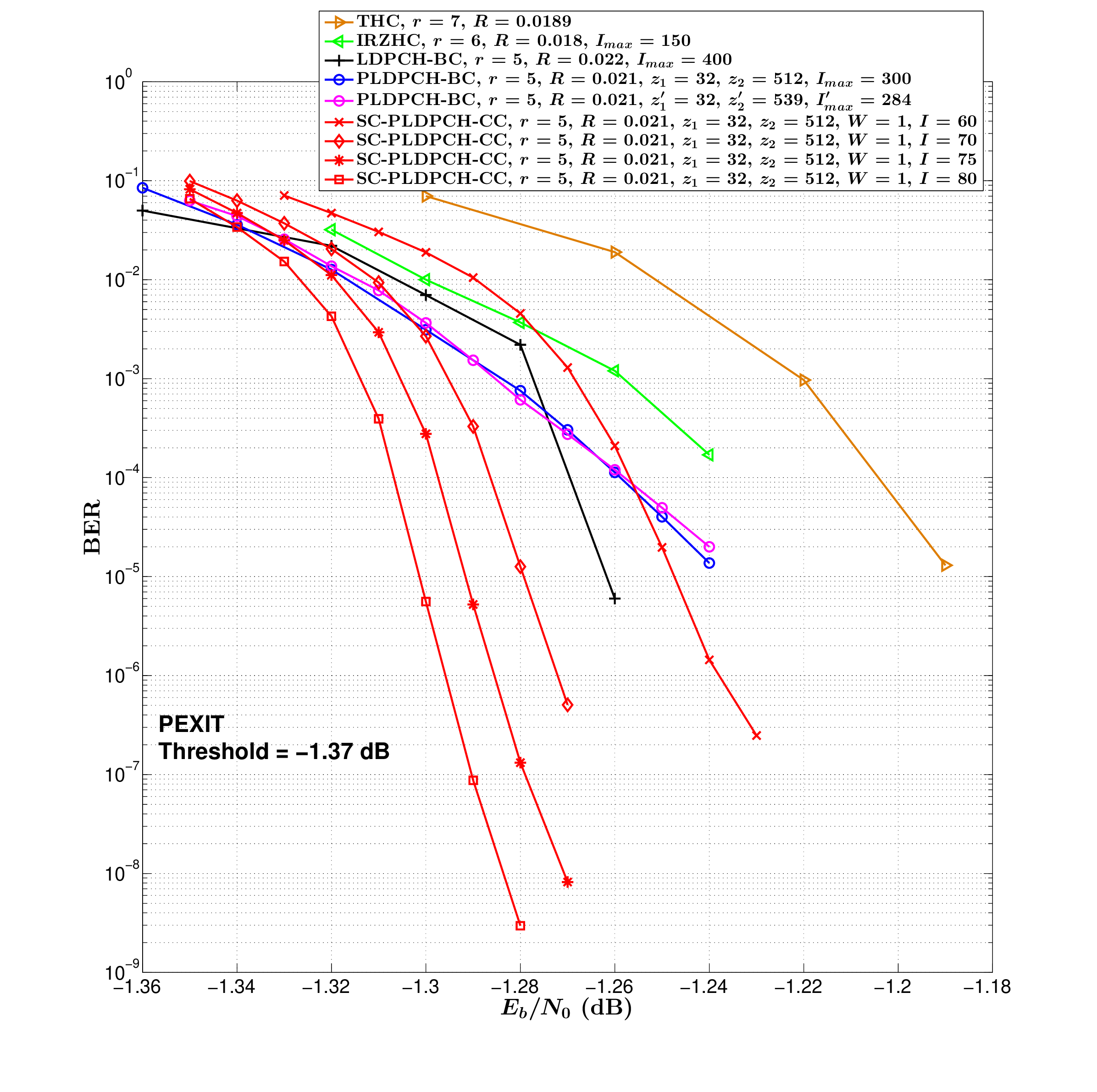}}
\caption{\color{black}BER performance of rate-$0.021$ SC-PLDPCH-CC
{\color{black}(PEXIT threshold = $-1.37$ dB)}, rate-$0.021$ PLDPCH-BC \cite{zhang2021}, rate-0.0189 THC \cite{Li2003}, rate-0.018 IRZHC \cite{Li2008b} and rate-0.022 LDPCH-BC \cite{Yue2007}.  Information lengths are around $65536$
 for all codes.  
 $r = 5$, lifting factors $z_1=32$ and $z_2 = 512$ for SC-PLDPCH-CC.
 \color{black} $r = 5$, lifting factor sets (i) $z_1=32$ and $z_2 = 512$; 
 and (ii) $z'_1=32$ and $z'_2 = 539$  for PLDPCH-BC.
}
  \label{fig:r5}
\end{figure}

\subsection{Rate-$0.008$ and $r = 8$, Rate-$0.00295$ and $r = 10$}
Based on the $5\times 15$ protomatrix $\bm{B}$ of the optimized rate-$0.008$ PLDPCH-BC \cite{zhang2021}, we apply our proposed GA and find the protomatrices
shown as \eqref{mat:r8_b0} and  \eqref{mat:r8_b1}.
The protomatrices are found after $59$ generations and the corresponding fitness value equals $457$.
We increase $L$ to $300$ such that the SC-PLDPCH-TDC has the same code rate with its underlying block code \cite{zhang2021}.
Using our proposed PEXIT method, the terminated code with $L=300$ has a threshold of $-1.45$ dB as shown in Table \ref{tb:threshold}.
Hence, the theoretical threshold of the SC-PLDPCH-CC constructed by \eqref{mat:r8_b0} and \eqref{mat:r8_b1} approximately equals $-1.45$ dB, which is slightly lower {\color{black}(i.e., slightly better)} than that of the PLDPCH-BC \cite{zhang2021}.
We lift the SC-PLDPCH-CC with factors $z_1=16$ and $z_2=1280$ such that
its sub-block length is the same as that of the PLDPCH-BC in \cite{zhang2021}.
{\color{black} Moreover, the information lengths in each sub-block/block equal $204800$ for both codes.}

Fig. \ref{fig:r8_r10}(a) shows the BER performance of the two codes.
{\color{black} An} SC-PLDPCH-CC pipeline decoder with $I=100$ processors
achieves a BER of $10^{-5}$ at about $E_b/N_0 = -1.40$ dB,
which outperforms that with (i) $I = 90$ by about $0.01$ dB,
(ii) $I = 80$ by about $0.025$ dB,
(iii) and $I = 75$ by about $0.035$ dB.
{\color{black} The gaps of the SC-PLDPCH-CC (with $I=100$ and BER of $10^{-5}$) to the  PEXIT threshold ($-1.45$ dB) and to the ultimate Shannon limit ($-1.59$) dB are $0.05$ dB and $0.19$ dB, respectively.
}
{\color{black}
We  can also observe that under the same latency, SC-PLDPCH-CC using $I = 75$ processors outperforms PLDPCH-BC using $I_{BC}=150$ layered decoding iterations
(equivalent to $I_{max} = 300$ standard decoding iterations, i.e., the blue curve with symbol ``$\circ$'')
by about  $0.01$ dB at a BER of $10^{-5}$.
}
{\color{black}
In Fig. \ref{fig:r8_r10}(a), we re-draw the BER curve of rate-$0.008$ LDPCH-BC
{\color{black}\cite[Fig.~13]{Yue2007}}
 which has an information length of
$238000$.
We see that at a BER of $10^{-5}$, the required $E_b/N_0$ for the rate-$0.008$ LDPCH-BC (with  {\color{black}$I_{max}=400$} iterations) is $-1.38$ dB and that for our rate-$0.008$ SC-PLDPCH-CC (pipeline decoder containing $100$ processors) is $-1.40$ dB.
}

\begin{figure}[!hp]
\centerline{\includegraphics[width=0.6\columnwidth]{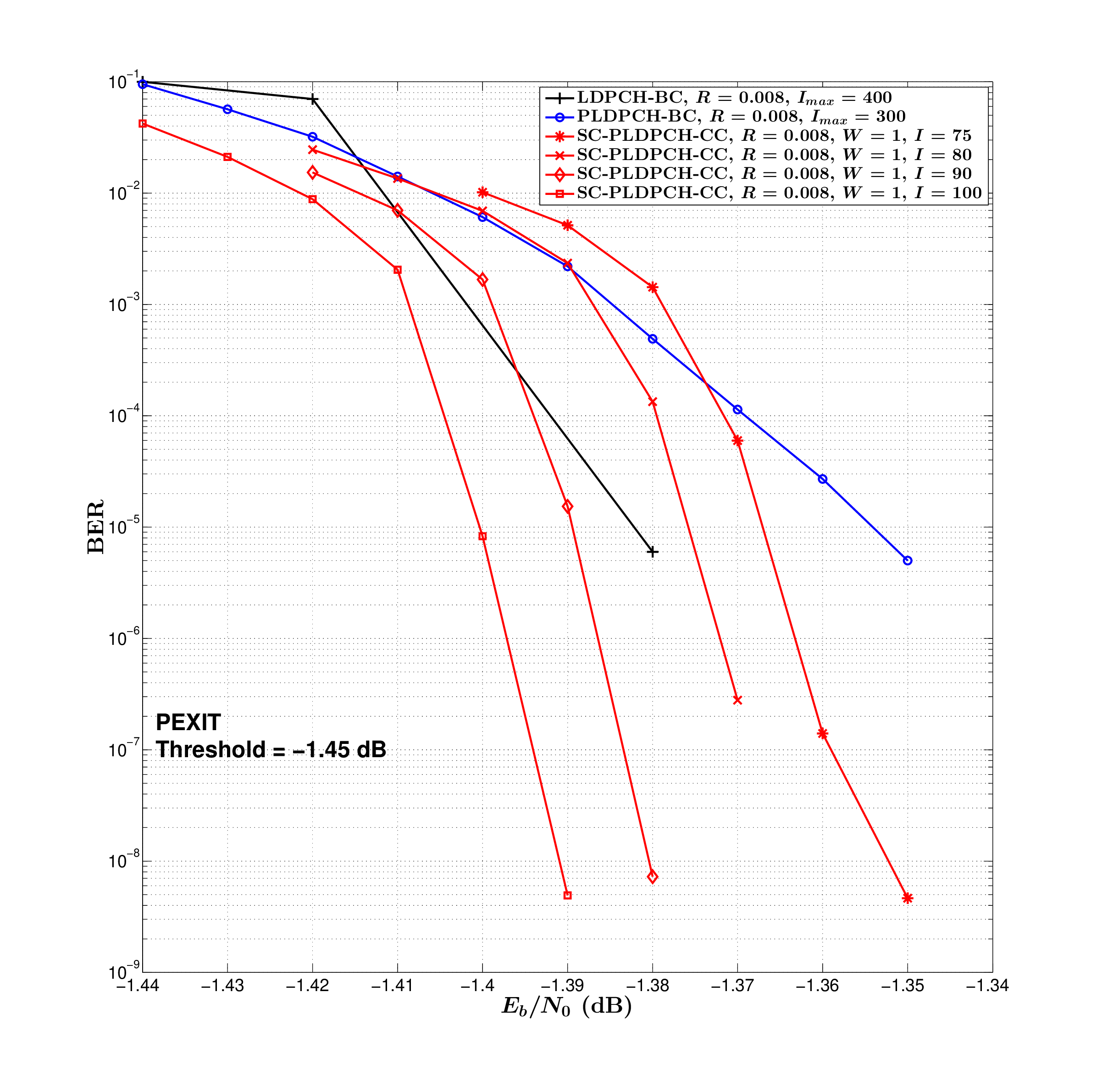}}
\centerline{ (a)} 
\centerline{\includegraphics[width=0.6\columnwidth]{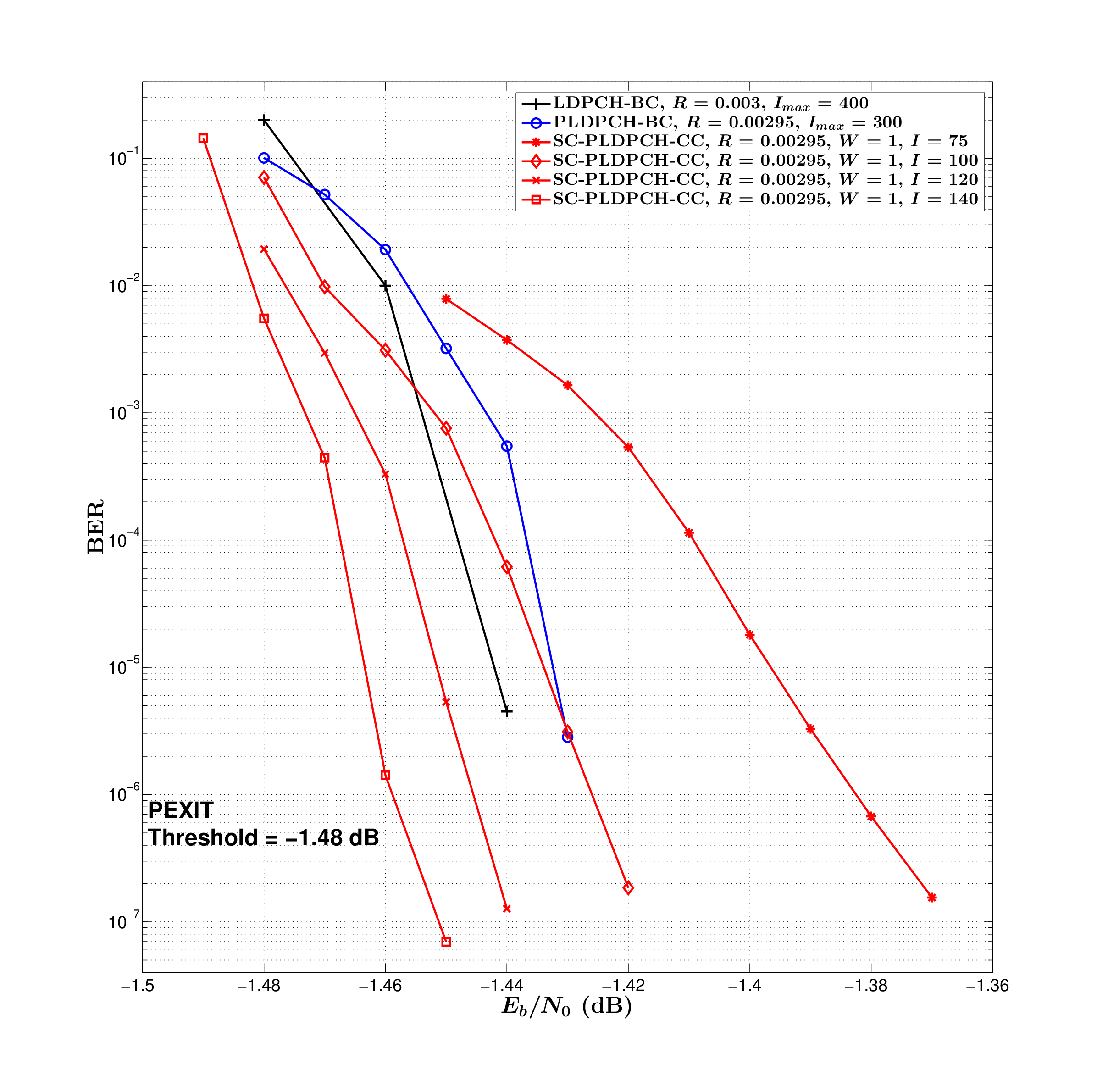}}
\centerline{  (b)}
\caption{\color{black}(a) BER performance of rate-$0.008$ SC-PLDPCH-CC
{\color{black} (PEXIT threshold = $-1.45$ dB)}, rate-$0.008$ PLDPCH-BC \cite{zhang2021}
and rate-0.008 LDPCH-BC \cite{Yue2007}.
Information length is $204,800$ for both SC-PLDPCH-CC and PLDPCH-BC;
and $238,000$ for LDPCH-BC. $r=8$, lifting factors $z_1=16$ and $z_2 = 1,280$ for both SC-PLDPCH-CC and PLDPCH-BC. (b) BER performance
of rate-$0.00295$ SC-PLDPCH-CC
{\color{black} (PEXIT threshold = $-1.48$ dB)}, rate-$0.00295$ PLDPCH-BC \cite{zhang2021} and
 rate-0.003 LDPCH-BC \cite{Yue2007}.
 Information length is $460,800$ for both SC-PLDPCH-CC and PLDPCH-BC;
and $650,00$ for LDPCH-BC.
 $r=10$, lifting factors $z_1=20$ and $z_2 = 1,280$ for both SC-PLDPCH-CC and PLDPCH-BC.
 }
  \label{fig:r8_r10}
\end{figure}

Based on the $6\times 24$ protomatrix $\bm{B}$ of the optimized rate-$0.00295$ PLDPCH-BC \cite{zhang2021}, the two protomatrices $\bm{B}_0$
and $\bm{B}_1$ obtained by our proposed GA are shown in
 \eqref{mat:r10_b0}  and \eqref{mat:r10_b1}, respectively.
The protomatrices \eqref{mat:r10_b0}  and \eqref{mat:r10_b1} are found after $48$ generations, and the corresponding fitness value equals $516$.
Using the proposed PEXIT method, Table \ref{tb:threshold} lists the thresholds for SC-PLDPCH-TDC with $L = 100$ and the underlying PLDPCH-BC \cite{zhang2021}.
The SC-PLDPCH-TDC constructed by $L=100$ sets of protomatrices \eqref{mat:r10_b0} and \eqref{mat:r10_b1} has almost the same code rate with its underlying block code, and its threshold is estimated to be $-1.48$ dB.
Hence, the theoretical threshold of the SC-PLDPCH-CC constructed by \eqref{mat:r10_b0} and \eqref{mat:r10_b1} also equals $-1.48$ dB, which is slightly greater than that of the PLDPCH-BC \cite{zhang2021}.
We lift the SC-PLDPCH-CC with factors $z_1=20$ and $z_2=1280$ such that
its sub-block length is the same as that of the PLDPCH-BC
 \cite{zhang2021}.
Fig. \ref{fig:r8_r10}(b) shows the BER performance of the two codes.
SC-PLDPCH-CC decoder with $I=140$ processors
achieves a BER of $10^{-5}$ at about $E_b/N_0 = -1.465$ dB,
which outperforms that with (i) $I = 120$ by about $0.01$ dB,
(ii) $I = 100$ by about $0.03$ dB,
and (iii)  $I = 75$ by about $0.07$ dB.
{\color{black} At a BER of $10^{-5}$, the gaps (for the SC-PLDPCH-CC with $I=140$ iterations) to the  PEXIT threshold ($-1.48$ dB) and to the ultimate Shannon limit ($-1.59$) dB are $0.015$ dB and $0.125$ dB, respectively.}
{\color{black}
Under the same latency, SC-PLDPCH-CC using $I = 75$ processors is outperformed by PLDPCH-BC using $I_{BC}=150$ layered decoding iterations
(equivalent to  {\color{black}$I_{max}=300$} standard decoding iterations) by about  $0.04$ dB at a BER of $10^{-5}$.
}

{\color{black}
In Fig. \ref{fig:r8_r10}(b), we redraw the BER curve of rate-$0.003$ LDPCH-BC
{\color{black}\cite[Fig.~13]{Yue2007}}
 which has an information length of $650,000$. Results show that the rate-$0.003$ LDPCH-BC (with  {\color{black}$I_{max}=400$} iterations) achieves a BER of $10^{-5}$ at $E_b/N_0 = -1.44$ dB and the rate-$0.00295$ SC-PLDPCH-CC (containing $I=140$ processors) achieves the same error performance at $E_b/N_0 = -1.465$ dB.
}

\section{Conclusion}\label{sect:conclusion}
We have proposed a new type of ultimate-Shannon-limit-approaching code
called 
SC-PLDPCH-CCs,
which
are formed by
 spatially coupling 
 PLDPCH-BCs.
We have developed a pipelined decoding with layered scheduling algorithm to efficiently and effectively decode SC-PLDPCH-CCs, and
 have proposed a layered PEXIT method to evaluate the threshold of 
 SC-PLDPCH-TDCs.
Based on the protomatrix of a PLDPCH-BC, we have proposed a genetic algorithm to systematically search for the protomatrices of good SC-PLDPCH-TDCs.
We extend the coupling length of these SC-PLDPCH-TDCs with good thresholds to form SC-PLDPCH-CCs.
Using the proposed methods, we have found SC-PLDPCH-CCs
which are comparable to or superior to their block code counterparts in terms of theoretical threshold and error performance.
{\color{black}Their simulated error performances outperform state-of-the-art low-rate codes with comparable rates and information lengths.
}
At a BER of $10^{-5}$, the SC-PLDPCH-CCs
of rates $0.0494$, $0.021$, $0.008$ and $0.00295$ are only
 $0.352$ dB, $0.29$ dB, $0.19$ dB and $0.125$ dB
from the ultimate Shannon limit. 


\bibliographystyle{ieeetr}
\bibliography{Reference}

\begin{IEEEbiography}
[{\includegraphics[width=1in,height=1.25in,clip,keepaspectratio]{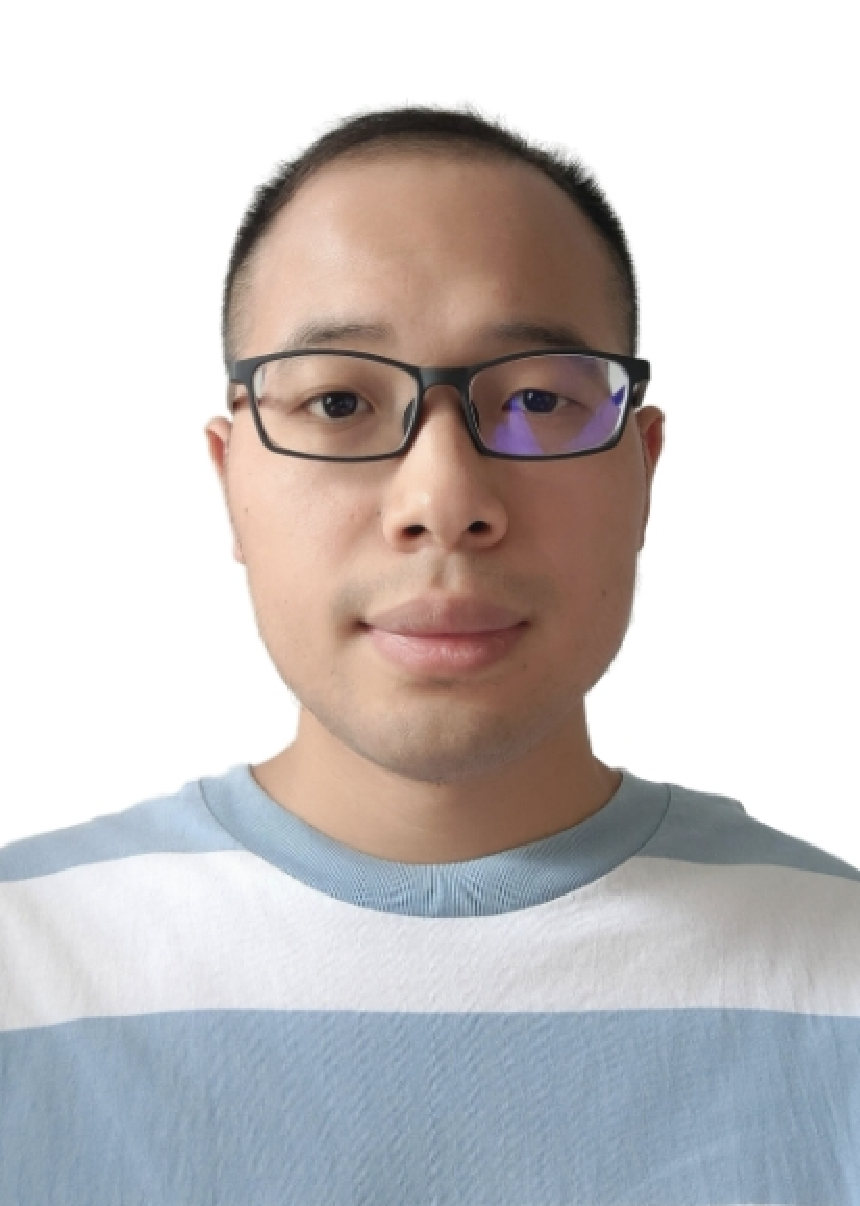}}]
{Peng-Wei Zhang} received the Bachelor of Engineering degree in Electronics and Information Engineering and
the Master of Engineering degree in Electronics and Communication Engineering from Chongqing University of Posts and Telecommunications, China, in 2013 and 2016, respectively.
He received his PhD degree at the Department of Electronic and Information Engineering, The Hong Kong Polytechnic University, Hong Kong SAR, China, in 2021. He is now with Huawei Technologies Ltd., Chengdu, China.
\end{IEEEbiography}

\begin{IEEEbiography}[{\includegraphics[width=1in,height=1.25in,clip,keepaspectratio]{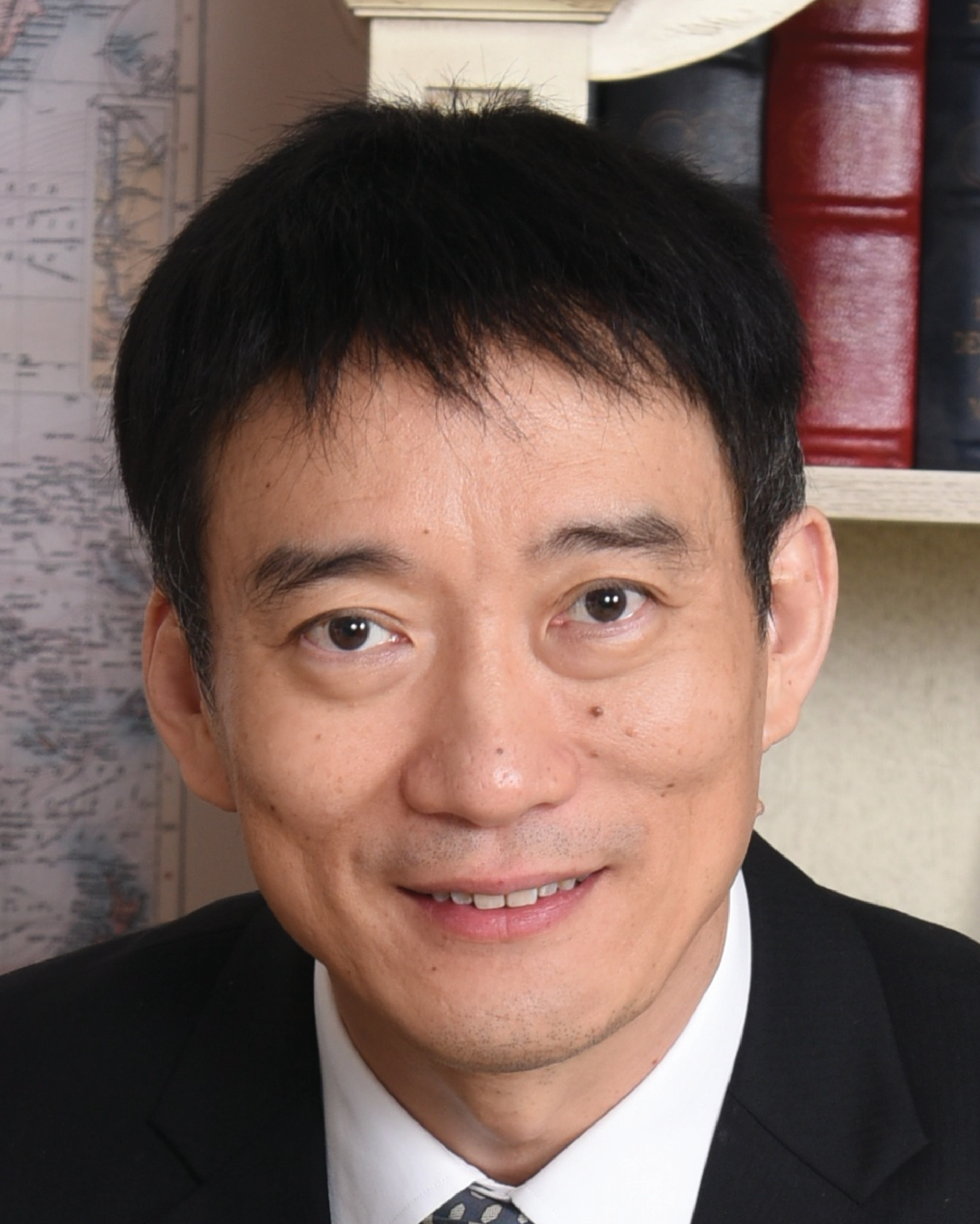}}]{Francis C. M. Lau} received the BEng(Hons) degree in electrical and electronic engineering and the PhD degree from King's College London, University of London, UK. He is a Professor at the Department of Electronic and Information Engineering, The Hong Kong Polytechnic University, Hong Kong. He is also a Fellow of IEEE and a Fellow of IET. 

He is a co-author of two research monographs. He is also a co-holder of six US patents. He has published more than 330 papers. His main research interests include channel coding, cooperative networks, wireless sensor networks, chaos-based digital communications, applications of complex-network theories, and wireless communications. 
He is a co-recipient of one Natural Science Award from the Guangdong Provincial Government, China; eight best/outstanding conference paper awards; one technology transfer award; two young scientist awards from International Union of Radio Science; and one FPGA design competition award.

He was the General Co-chair of International Symposium on Turbo Codes \& Iterative Information Processing (2018) and the Chair of Technical Committee on Nonlinear Circuits and Systems, IEEE Circuits and Systems Society (2012-13). He served as an associate editor for IEEE TRANSACTIONS ON CIRCUITS AND SYSTEMS II (2004-2005 and 2015-2019),  IEEE TRANSACTIONS ON CIRCUITS AND SYSTEMS I (2006-2007), and IEEE CIRCUITS AND SYSTEMS MAGAZINE (2012-2015). He has been a guest associate editor of INTERNATIONAL JOURNAL AND BIFURCATION AND CHAOS since 2010. 
\end{IEEEbiography}

\begin{IEEEbiography}
[{\includegraphics[width=1in,height=1.25in,clip,keepaspectratio]
{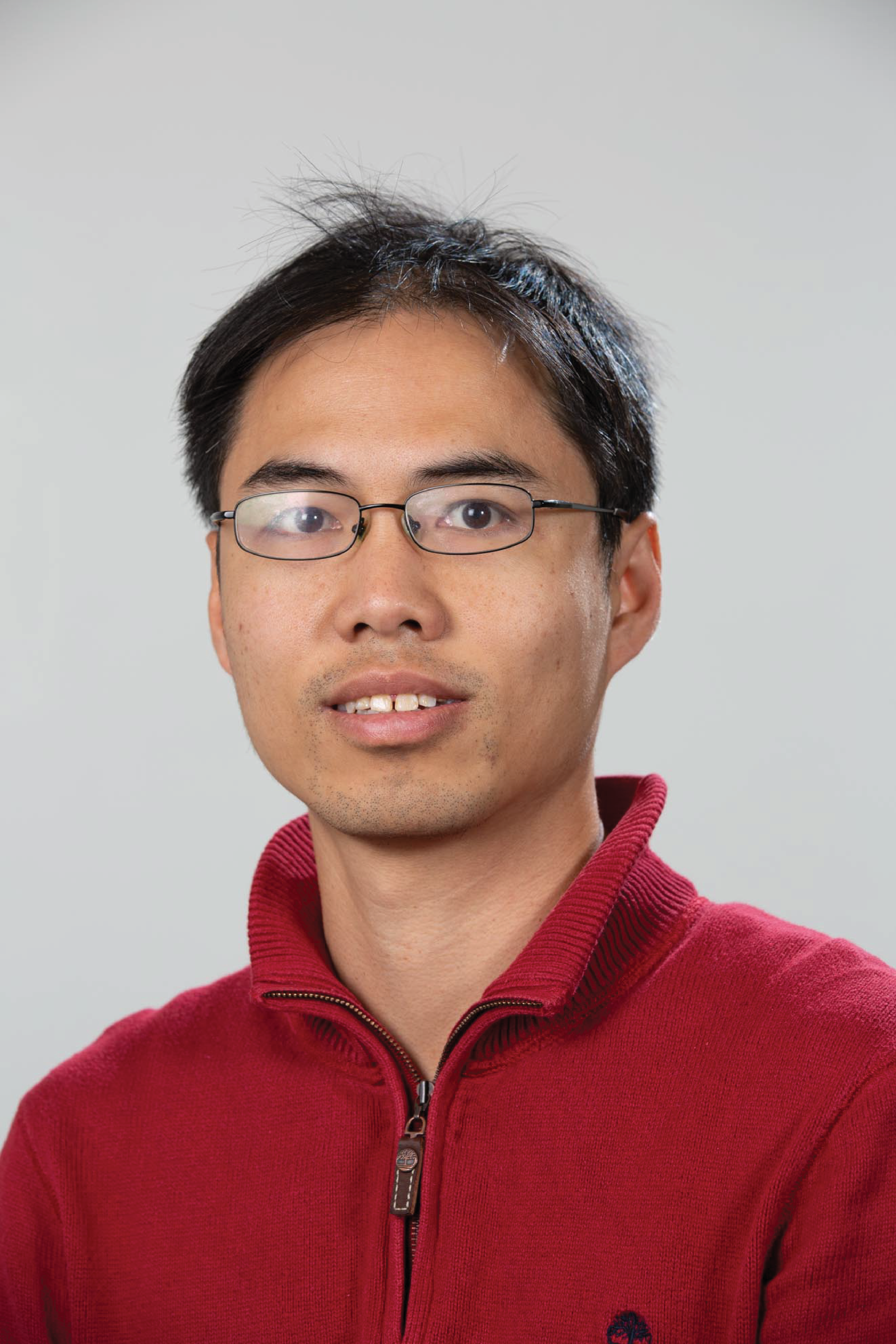}}]
{Chiu-Wing Sham} received his Bachelor degree (Computer Engineering) and  MPhil. degree from The Chinese University of Hong Kong in 2000 and 2002 respectively, and received his Ph.D. degree from the same university in 2006. He has worked as an Electronic Engineer on the FPGA applications of the motion-control system and system security with cryptography in ASM Pacific Technology Ltd (HK). During the years at The Hong Kong Polytechnic University, he engaged in various University projects for the commercialization of technology, in particular, a few optical communication projects which were in collaboration with Huawei. He also worked on the physical design of VLSI design automation. He was invited to work at Synopsys, Inc. (Shanghai) in the summer of 2005 as a Visiting Research Engineer. He is now working at The University of Auckland as a Senior Lecturer. He is also an IEEE Senior Member and an Associate Editor of IEEE TRANSACTIONS ON CIRCUITS AND SYSTEMS II (2017-present).
\end{IEEEbiography}

\end{document}